\documentclass[prd,preprintnumbers,superscriptaddress,nofootinbib,amsmath,amsfonts,amssymb]{revtex4}
\usepackage[a4paper, hdivide={1.91cm,,1.165cm}, vdivide={1.83cm,,3.6cm}]{geometry}
\usepackage{graphicx}
\usepackage{hyperref}
\usepackage[dvipsnames]{xcolor}
\hypersetup{colorlinks=true,linkcolor=Maroon,citecolor=ForestGreen,filecolor=ForestGreen,urlcolor=ForestGreen}
\DeclareUnicodeCharacter{2212}{-}

\usepackage{soul}

\usepackage{caption}
\usepackage{subcaption}

\def\bpm{\begin{pmatrix}}
\def\epm{\end{pmatrix}}

\begin{document}
\title{Leptonic probes of Alternative Left-Right Symmetric Models}

\author{Mariana Frank}
\email{mariana.frank@concordia.ca}
\affiliation{Department of Physics, Concordia University,7141 Sherbrooke St. West, Montreal, QC H4B 1R6, Canada}

\author{Benjamin Fuks}
\email{fuks@lpthe.jussieu.fr}
\affiliation{Laboratoire de Physique Th\'{e}orique et Hautes \'{E}nergies (LPTHE), UMR 7589,\\ Sorbonne Universit\'{e} \& CNRS, 4 place Jussieu, 75252 Paris Cedex 05, France}
\affiliation{Department of Physics, Indian Institute of Technology Guwahati, Assam 781 039, India}

\author{Sumit K. Garg}
\email{sumit.kumar@manipal.edu}
\affiliation{Manipal Centre for Natural Sciences, Manipal Academy of Higher Education, Dr.T.M.A. Pai Planetarium Building, Manipal-576104, Karnataka, India}

\author{Chayan~Majumdar}
\email{c.majumdar@ucl.ac.uk}
\affiliation{Department of Physics and Astronomy, University College London, London WC1E 6BT, United Kingdom}

\author{Poulose Poulose}
\email{poulose@iitg.ac.in}
\affiliation{Department of Physics, Indian Institute of Technology Guwahati, Assam 781 039, India}

\author{Supriya Senapati}
\email{ssenapati@umass.edu}
\affiliation{Amherst Center for Fundamental Interactions, Department of Physics, University of Massachusetts, Amherst, MA 01003, USA}

\begin{abstract}
We explore constraints on the parameter space of the alternative left-right model originating from the leptonic sector. Our analyses focuses on both lepton-flavour-conserving observables, particularly the anomalous magnetic moment of the muon, and lepton-flavour-violating processes like $\mu \to e \gamma $ decay and $\mu-e$ conversions in nuclei. While contributions to the anomalous magnetic moment fall below the measured values at 2$\sigma$, current and future experimental sensitivities to flavour-violating branching rations of the Standard Model leptons are expected to impose lower bounds on the mass of the peculiar $SU(2)_R$ gauge boson of the model. This provides complementary constraints relative to existing limits, which are indirect and derived from collider bounds on the mass of the associated neutral gauge boson $Z^\prime$.
\end{abstract}

\pacs{}
\maketitle

\section{Introduction}
\label{sec:intro}
The Standard Model (SM) of particle physics, based on the gauge group $SU(3)_C \otimes SU(2)_L \otimes U(1)_Y$, successfully describes the properties of the elementary particles and their interactions. While its predictions match experimental data with remarkable accuracy, the SM faces several unresolved puzzles and conceptual challenges. Notably, it is a chiral framework, with maximal parity violation through the weak interaction, while the strong and electromagnetic interactions conserve parity. Additionally, the SM does not include right-handed neutrinos or extra fermionic states, preventing it from explaining observations from neutrino experiments which have firmly established that neutrinos are massive and they mix. Among the various theoretical frameworks proposed to extend the SM and address these issues, left-right symmetric models (LRSMs)~\cite{Pati:1974yy, Mohapatra:1974gc, Senjanovic:1975rk, Mohapatra:1977mj} offer an attractive alternative. Based on the gauge group $SU(3)_C \otimes SU(2)_L \otimes SU(2)_R \otimes U(1)_{B-L}$, LRSMs relate the maximal breaking of parity to the smallness of the neutrino masses. By postulating parity invariance of weak interactions via interchangeable $SU(2)_L$ and $SU(2)_R$ symmetries, such frameworks automatically include three right-handed neutrinos needed  for the implementation of seesaw mechanism to explain the smallness of the observed neutrino masses. Moreover, the $U(1)$ charge assignments are less arbitrary than in the SM, as the hypercharge quantum numbers now emerge from the mixing of the (gauged) $B-L$ symmetry and $SU(2)_R$, both having a physical origin. 

The left-right symmetry can also be seen as emerging from a grand-unified framework based on the exceptional group $E_6$~\cite{Gursey:1975ki, Achiman:1978vg, Hewett:1988xc}, which is broken to its maximal subgroup $SU(3)_C \otimes SU(3)_L \otimes SU(3)_H$. The first of these $SU(3)$ factors remains unbroken and is identified with the SM strong interaction symmetry, while the other two are further broken down to $SU(2)_L \otimes SU(2)_H \otimes U(1)_X$ and then to the electroweak symmetry group. According to this structure, the fermions of the theory lie in $SU(3)_H$ triplets. There are three different ways to form an $SU(2)_H$ doublet from the three components of an $SU(3)_H$ triplet~\cite{Ma:1986we}. Canonical LRSM constructions identify $SU(2)_H$ with $SU(2)_R$, and impose that the first two components of the $SU(3)_H$ triplets are paired into $SU(2)_R$ doublets comprised of the right-handed SM fermions and neutrinos $\nu_R$ (which are singlets under the SM gauge symmetry). The symmetry breaking pattern requires an extended Higgs sector that features several scalar fields whose couplings to $SU(2)_L$ and $SU(2)_R$ doublets could yield unacceptably large tree-level flavour-changing neutral interactions. Consequently, conflicts with observations can only be avoided if the breaking of the $SU(2)_R$ symmetry occurs at a very high scale. As a result, the spectrum of new particles is pushed beyond the reach of current collider experiments, and the effects of the new states could only be observed indirectly through the study of rare phenomena.

Alternatively, the $SU(2)_H$ symmetry could be identified with another symmetry group, $SU(2)_{R'}$, such that the first and third components of $SU(3)_H$ triplets are paired into doublets. In contrast to conventional LRSM constructions, Alternative Left-Right Models (ALRMs)~\cite{Ma:1986we, Babu:1987kp, Frank:2005rb, Ma:2010us, Ashry:2013loa} impose the down-type quark $d_R$ and neutrino $\nu_R$ fields to be $SU(2)_{R'}$ singlets, while the up-type quark $u_R$ and charged lepton $e_R$ fields are paired into $SU(2)_{R'}$ doublets with an exotic quark field $d'_R$ and scotino field $n_R$, respectively. For parity invariance, the matter sector of the theory additionally includes exotic $SU(2)_L \otimes SU(2)_{R'}$ singlets, specifically the $d'_L$ and $n_L$ states ~\cite{Frank:2004vg}. These exotic matter states further allow the theory to include a phenomenologically viable dark matter candidate~\cite{Frank:2019nid, Frank:2022tbm}, and naturally prevent it from yielding large flavour-changing neutral currents, even for a left-right symmetry breaking scale of a few TeV. This originates from the structure of the couplings of the $SU(2)_{R'}$ charged gauge boson $W_R$, which always interacts with a pair of fermions comprising one SM fermion and one exotic, heavier, fermion. Consequently, the $W_R$ boson can neither be be produced from, nor decay into, a pair of ordinary quarks (or leptons), and its mass is thus free from any constraints that could be imposed from direct searches for extra gauge bosons at the LHC. Moreover, this boson does not mix with the charged SM $W$ ($\equiv W_L$) gauge boson. Consequently, the $W_R$ boson can be much lighter than in conventional LRSM scenarios. However indirect bounds on the $W_R$ mass exists through the relation between its mass and that of the additional neutral gauge boson $Z^\prime$, which can be produced and decay in the ordinary way and that is also allowed to mix with the neutral SM $Z$ boson. The most stringent bounds on the mass of the ALRM $W_R$ boson consequently arise indirectly from the bounds that are imposed on the mass of the neutral $Z^\prime$ boson \cite{ATLAS:2019erb,CMS:2021ctt}, through the relation between the masses of the two new bosons~\cite{Frank:2019nid}.

Furthermore, previous studies have established conditions for the stability of the ALRM vacuum state~\cite{Frank:2021ekj}, demonstrated the model's potential impact on neutrinoless double beta decay processes and leptogenesis~\cite{Frank:2020odd}, and shown that exotic fermions and Higgs bosons could indirectly influence rare top decays~\cite{Frank:2023fkc}. Notably, ALRM setup offers the possibility of significant enhancements in the corresponding branching ratios relative to the SM predictions, with predicted values close to the current limits. Building upon these last findings, we explore the implications of ALRM for rare lepton-flavour-violating (LFV) processes and on the anomalous magnetic moment of the muon $a_\mu$, and we investigate the extent to which this could yield potential constraints on the masses in the model. Loop effects are expected to be sensitive to contributions involving $W_R$-boson exchanges, so bounds on the considered processes could lead to $W_R$ mass limits stronger than the indirect collider bounds derived from $Z^\prime$ constraints. This would highlight an interesting complementarity between direct searches for extra gauge bosons at colliders and indirect bounds from rare LFV processes and $a_\mu$. While collider mass limit on ALRM $Z^\prime$ bosons lead to indirect bounds on the $W_R$ state, the non-observation of rare LFV processes and the expected size of the new physics contributions to $a_\mu$ would constrain not only the $W_R$ boson itself, but also the $Z^\prime$ state indirectly, via the corresponding mass relations.

The structure of the paper is as follows. In Section~\ref{sec:mod}, we briefly review the field content of the ALRM, the associated Lagrangian and  relevant relations between the physical masses and the  free parameters in the model. Section~\ref{sec:gm2} is dedicated to a study of the impact of the one-loop and two-loop ALRM contributions to the anomalous magnetic moment of the muon and the extraction of associated constraints, while in Section~\ref{sec:LFV} we focus on rare LFV processes such as $\mu\to e\gamma$ and $\mu-e$ conversions in various nuclei, and compare predictions to present and future bounds. We present our conclusions in Section~\ref{sec:conclusion}.

\section{Description of the ALRM Framework} \label{sec:mod}
\begin{table}
 \centering
  \setlength\tabcolsep{1.5pt}
  \begin{tabular}{ccc}
    Field & $\mathcal{G}_\mathrm{ALRM}$ &$U(1)_S$\\\hline
    $\chi_L = \bpm\chi_L^+\\\chi_L^0\epm$ & $\big({\bf 1}, {\bf 2}, {\bf 1}\big)_{\tfrac12}$ & \phantom{-}\ 0\\
    $\chi_R = \bpm\chi_R^+\\\chi_R^0\epm$ & $\big({\bf 1}, {\bf 1}, {\bf 2}\big)_{\tfrac12}$ &$\phantom{-}\tfrac12$\\
    $\Phi = \bpm \phi_1^0&\phi_2^+\\ \phi_1^-&\phi^0_2 \epm$ & $\big({\bf 1}, {\bf 2}, {\bf 2}^*\big)_{0}$ & $-\tfrac12$\\[.5cm]
  \end{tabular}\hspace{.5cm}
  \begin{tabular}{ccc}
    Field & $\mathcal{G}_\mathrm{ALRM}$ &$U(1)_S$\\\hline
    $Q_L = \bpm u_L\\d_L\epm$ & $\big({\bf 3}, {\bf 2}, {\bf 1}\big)_{\tfrac16}$ & \phantom{-}\ 0\\
    $Q_R = \bpm u_R\\d_R'\epm$ & $\big({\bf 3}, {\bf 1}, {\bf 2}\big)_{\tfrac16}$ & $-\tfrac12$\\
    $d'_L$ & $\big({\bf 3}, {\bf 1}, {\bf 1}\big)_{-\tfrac13}$ & $-1$\\
    $d_R$  & $\big({\bf 3}, {\bf 1}, {\bf 1}\big)_{-\tfrac13}$ & \phantom{-}\ 0\\
  \end{tabular}  \hspace{.4cm}
  \begin{tabular}{ccc}
    Field & $\mathcal{G}_\mathrm{ALRM}$ &$U(1)_S$\\\hline
    $L_L = \bpm \nu_L\\e_L\epm$ & $\big({\bf 1}, {\bf 2}, {\bf 1}\big)_{-\tfrac12}$ & 1 \\
    $L_R = \bpm n_R\\e_R\epm$  & $\big({\bf 1}, {\bf 1}, {\bf 2}\big)_{-\tfrac12}$ & $\tfrac32$ \\
    $n_L$   & $\big({\bf 1}, {\bf 1}, {\bf 1}\big)_{0}$ & 2 \\
    $\nu_R$ & $\big({\bf 1}, {\bf 1}, {\bf 1}\big)_{0}$ & 1 \\[.1cm]
  \end{tabular} 
  \caption{ALRM scalar (left), quark (centre) and lepton (right) fields, their representation under $\mathcal{G}_\mathrm{ALRM}$, and their $U(1)_S$ charge assignments.\label{tab_ALRM}}
\end{table}

The ALRM relies on the $\mathcal{G}_\mathrm{ALRM} = SU(3)_C \otimes SU(2)_L \otimes SU(2)_{R'} \otimes U(1)_{B-L}$ gauge symmetry, supplemented by an additional $U(1)_S$ global symmetry. The generator of this global symmetry, along with the third generator of $SU(2)_{R'}$, allows for the definition of a conserved generalised lepton number and dark matter stability. The spontaneous breaking of $SU(2)_{R'} \otimes U(1)_{B-L} \otimes U(1)_S$ to the hypercharge group is implemented by means of an $SU(2)_{R'}$ doublet of Higgs fields $\chi_R$ charged under $U(1)_{B-L}$, while the breaking of $SU(2)_L\otimes U(1)_Y$ to electromagnetism proceeds via a bi-doublet of Higgs fields charged under both $SU(2)_L$ and $SU(2)_{R'}$, with zero $B-L$ charge. Furthermore, to maintain left-right symmetry, the scalar sector of the model includes an $SU(2)_L$ doublet $\chi_L$. The complete field content of the ALRM scalar sector, along with the quantum numbers under $\mathcal{G}_{ALRM}\otimes U(1)_S$, is provided in the left panel of Table~\ref{tab_ALRM}. The structure of the vacuum corresponding to the breaking pattern mentioned above is given by
\begin{equation}
    \langle \Phi \rangle = \tfrac{1}{\sqrt{2}} \bpm  0 & 0 \\ 0 & k\epm, \qquad
    \langle \chi_L \rangle = \tfrac{1}{\sqrt{2}} \begin{pmatrix} 0  \\ v_L \end{pmatrix},\qquad
    \langle \chi_R \rangle = \tfrac{1}{\sqrt{2}} \begin{pmatrix} 0 \\ v_R \end{pmatrix}\,,
\end{equation}
where the component field $\phi^0_1$ is protected from acquiring a vacuum expectation value (VEV) by the conservation of the generalised lepton number.

The central and right panels of Table~\ref{tab_ALRM} present the fermionic field content of the model. It includes the conventional $SU(2)_L$ doublets of left-handed quarks and leptons, $Q_L$ and $L_L$, as well as the $SU(2)_L$ singlets of right-handed down-type quarks $d_R$ and neutrinos $\nu_R$. All these fields are also $SU(2)_{R'}$ singlets. Additionally, the model features two $SU(2)_{R'}$ doublets that pair the SM right-handed quark $u_R$ and charged lepton $e_R$ with two exotic fermions, a right-handed down-type quark $d'_R$ and a scotino $n_R$, respectively. Moreover, to preserve the left-right symmetry, the model also contains two exotic $SU(2)_{R'}$ singlets, a new left-handed down-type quark $d'_L$ and a left-handed scotino $n_L$. All fermionic fields become massive after the breaking of $\mathcal{G}_\mathrm{ALRM}$ to electromagnetism, and those with the same quantum numbers mix. However, the conservation of the generalised lepton number forbids scotino/neutrino and SM/exotic down-type quark mixings, and also prevents the $SU(2)_L$ and $SU(2)_{R'}$ charged gauge bosons $W\equiv W_L$ and $W_R$ from mixing.

The model Lagrangian includes standard gauge-invariant kinetic terms for all fields, as well as Yukawa interactions between the Higgs sector and the fermions,
\begin{equation}\begin{split}
    \mathcal{L}_Y = &\
      -\bar Q_L { Y}^q \tilde\Phi^\dag Q_R
    - \bar Q_L {Y}^q_L \chi_L d_R
    - \bar Q_R { Y}^{q}_R \chi_R d'_L
    - \bar L_L { Y}^\ell \Phi L_R
    - \bar L_L { Y}^\ell_L \tilde \chi_L^\dag \nu_R
    - \bar L_R { Y}^\ell_R \tilde \chi_R^\dag n_L 
    + \bar \nu_{R}^c M \nu_{R}
    + \mathrm{H.c.}
\end{split}\end{equation}
In our notation, the Yukawa couplings ${Y}$ are $3 \times 3$ matrices in the flavour space (with generation indices not shown for clarity.
The soft $U(1)_S$ symmetry breaking term,
 $\bar \nu_{R}^c M \nu_{R}$ (with $M$ being also a $3\times 3$  matrix in the flavour space),  is introduced to generate neutrino masses via  seesaw mechanism. This implies relations between the neutrino and scotino mass eigenstates $\nu$, $N$, $\hat{n}$ and their flavour-eigenstate counterparts via five $3\times3$ mixing matrices $\mathcal{V}^{\nu \nu}$, $\mathcal{V}^{\nu N}$, $\mathcal{V}^{N \nu}$, $\mathcal{V}^{NN}$ and $\mathcal{V}^{nn}$,
\begin{equation}
\nu_{\alpha L} = {\cal V}^{\nu\nu}_{\alpha i}\nu_i+ {\cal V}^{\nu N}_{\alpha j} N_j,~~~~~~
\nu_{Ri} = {\cal V}^{N \nu}_{ ij}\nu_j+ {\cal V}^{N N}_{i j} N_j,~~~~~~~
n_{Ri} =  {\cal V}^{nn}_{ij}\hat n_{Rj}
\end{equation}
where $\alpha = e, \mu, \tau$ and $(i,j)=1,2,3$. Furthermore, we have absorbed all scotino rotations into a redefinition of the right-handed scotino basis $\{n_R\} \to \{\hat{n}_R\}$, the $\{n_L\} = \{\hat{n}_L\}$ field basis being unchanged.

The multiscalar interactions are encoded in the scalar potential
\begin{equation}\begin{split}
  \hspace*{-.3cm}V = &\
    \kappa \big[\chi_L^\dag \Phi \chi_R \!+\! \chi_R^\dag\Phi^\dag\chi_L\big]
   - \mu_1^2 {\rm Tr} \big[\Phi^\dag \Phi\big]
   - \mu_2^2 \big[\chi_L^\dag \chi_L \!+\! \chi_R^\dag \chi_R\big]
   + \lambda_1 \big({\rm Tr}\big[\Phi^\dag \Phi\big]\big)^2
   + \lambda_2\ (\Phi\!\cdot\!\tilde\Phi) (\tilde\Phi^\dag\!\cdot\!\Phi^\dag)\\&
   + \lambda_3 \Big[\big(\chi_L^\dag \chi_L\big)^2 + \big(\chi_R^\dag\chi_R\big)^2\Big]
   + 2 \lambda_4\ \big(\chi_L^\dag \chi_L\big)\ \big(\chi_R^\dag\chi_R\big)
   + 2 \alpha_1 {\rm Tr} \big[\Phi^\dag \Phi\big] \big[\chi_L^\dag \chi_L \!+\! \chi_R^\dag \chi_R\big] \\&
   + 2 \alpha_2 \big[ \big(\chi_L^\dag \Phi\big) \big(\chi_L\Phi^\dag\big) +
          \big(\Phi^\dag \chi_R^\dag\big)\ \big(\Phi\chi_R\big)\big]
   + 2 \alpha_3 \big[ \big(\chi_L^\dag \tilde\Phi^\dag\big)\
          \big(\chi_L\tilde\Phi\big) + \big(\tilde\Phi\chi_R^\dag\big)\
          \big(\tilde\Phi^\dag\chi_R\big)\big]\,,
\end{split}\end{equation}
that includes bilinear ($\mu$), trilinear ($\kappa$) and quartic ($\lambda$, $\alpha$) terms. After the breaking of the $\mathcal{G}_\mathrm{ALRM}\otimes U(1)_S$ symmetry, all scalar fields with the same electric charge mix, with the exception of $\phi_1^0$ which is protected from mixing by the conservation of the generalised lepton number. The physical states thus contain four $CP$-even Higgs bosons $H_1^0$, $H_2^0$, $H_3^0$ and the SM Higgs $h$, two $CP$-odd Higgs bosons $A_1^0$ and $A_2^0$, and two pairs of charged Higgs bosons $H_1^\pm$, $H_2^\pm$. Additionally the set of mixed eigenstates includes two neutral and two charged massless Goldstone bosons $G_1^0$, $G_2^0$, $G_1^\pm$ and $G_2^\pm$, which are absorbed by the neutral and charged gauge bosons as their longitudinal degrees of freedom. All these states are related to the various gauge eigenstates of Table~\ref{tab_ALRM} via unitary rotations,
\begin{equation}\begin{split}
    \Im\{\phi_1^0\}= A_1^0\,,\quad
    \Re\{\phi_1^0\}= H_1^0\,, \quad
    \begin{pmatrix} \Im\{\phi_2^0\} \\ \Im\{\chi_L^0\} \\ \Im\{\chi_R^0\}  \end{pmatrix} 
       = U_{3\times 3}^{\rm A} \begin{pmatrix}  A_2^0 \\ G_1^0 \\  G_2^0 \end{pmatrix}\,,\quad
    \begin{pmatrix} \Re\{\phi_2^0\} \\ \Re\{\chi_L^0\} \\ \Re\{\chi_R^0\} \end{pmatrix}
       = U_{3\times 3}^{\rm H} \begin{pmatrix} h \\ H_2^0 \\ H_3^0 \end{pmatrix}\,,\\[.2cm]
    \begin{pmatrix} \phi^\pm_2 \\ \chi_L^\pm \end{pmatrix}
      = \begin{pmatrix} \cos \beta & \sin \beta \\ -\sin \beta & \cos \beta \end{pmatrix}
        \begin{pmatrix} H_1^\pm \\ G_1^\pm \end{pmatrix}\,,\qquad\qquad\qquad
    \begin{pmatrix} \phi^\pm_1 \\ \chi_R^\pm \end{pmatrix} 
      = \begin{pmatrix} \cos \zeta & \sin \zeta \\ -\sin \zeta & \cos \zeta \end{pmatrix}
        \begin{pmatrix} H_2^\pm \\ G_2^\pm \end{pmatrix}\,.\hspace{0.7cm}
\end{split}\end{equation}
Analytical expressions for the two $3\times3$ mixing matrices $U_{3\times 3}^{\rm A}$ and $U_{3\times 3}^{\rm H}$ in terms of the model's free parameters can be found in \cite{Frank:2019nid}, while the two mixing angles $\beta$ and $\zeta$ are defined by 
\begin{equation}
\tan \beta = k/v_L, \quad {\rm and} \quad \tan \zeta = k/v_R\,.
\end{equation}

The breaking of the left-right symmetry also generates masses for the model gauge bosons, and induces their mixing. The charged $W\equiv W_L$ and $W'\equiv W_R$ bosons do not mix, as $\langle\phi_1^0 \rangle = 0$, and their masses are given by
\begin{equation}
  M_{W_L}    = \frac12 g_L \sqrt{k^2+v_L^2} \equiv \frac12 g_L v
 \qquad {\rm and}\qquad
  M_{W_R} = \frac12 g_R \sqrt{k^2+v_R^2} \equiv \frac12 g_R v' \ ,
\label{eq:mw_mwp}\end{equation}
with $g_L$ and $g_R$ standing for the $SU(2)_L$ and $SU(2)_{R'}$ gauge coupling constants. In contrast, in the neutral sector, all states mix. The corresponding gauge boson squared mass matrix reads, in the $(B_\mu, W_{L\mu}^3, W_{R\mu}^3)$ basis,
\begin{equation}
  ({\cal M}^0_V)^2 = \frac14 \left(\begin{array}{ccc}
    g_{B-L}^2\ (v_L^2+v_R^2)  & -g_{B-L}\ g_L\ v_L^2    & -g_{B-L}\ g_R\ v_R^2\\
   -g_{B-L}\ g_L\ v_L^2       &  g_L^2\ v^2          & -g_L\ g_R\ k^2\\
   -g_{B-L}\ g_R\ v_R^2       & -g_L\ g_R\ k^2       &  g_R^2\ v^{\prime 2}
  \end{array} \right) ,
\end{equation}
where $g_{B-L}$ refers to the $B-L$ gauge coupling constant. This matrix can be diagonalised through three independent rotations that mix the $B$, $W_L^3$ and $W_R^3$ bosons into the massless photon $A$ and massive $Z$ and $Z^\prime$ states,
\renewcommand{\arraystretch}{1.}
\begin{equation}
 \left(  \begin{array} {c} B_\mu\\ W_{L\mu}^3\\ W_{R\mu}^3\end{array} \right)= 
\left ( \begin{array} {ccc}\cos \phi_W  & 0 & -\sin \phi_W \\ 0 & 1 & 0\\ \sin \phi_W & 0 & \cos \phi_W \end{array} \right) 
\left ( \begin{array} {ccc}\cos \theta_W & -\sin \theta_W & 0\\ \sin \theta_W & \cos \theta_W & 0\\ 0 & 0 & 1 \end{array} \right)
\left ( \begin{array} {ccc} 1 & 0 & 0\\ 0 & \cos \zeta_W & -\sin \zeta_W\\ 0 & \sin \zeta_W & \cos \zeta_W \end{array} \right)
 \left(  \begin{array} {c} A_\mu\\ Z_\mu\\ Z^\prime_{\mu} \end{array} \right) \,.
\end{equation}
The $\phi_W $-rotation mixes the $B$ and $W_R^3$ bosons into the hypercharge boson $B'$, the $\theta_W$-rotation corresponds to the usual electroweak mixing, and the $\zeta_W$-rotation is related to the strongly constrained $Z$/$Z^\prime$
mixing. The various mixing angles are related to the different gauge coupling constants and VEVs,
\begin{equation}\begin{split}
 & \sin \phi_W  = \frac{g_{B-L}}{\sqrt{g_{B-L}^2+g_R^2}} = \frac{g_Y}{g_R}
   \qquad\text{and}\qquad
   \sin \theta_W  = \frac{g_Y}{\sqrt{g_L^2+g_Y^2}} = \frac{e}{g_L} \ ,
 \\&
  \tan(2 \zeta_W)=\frac{2 \cos \phi_W  \cos \theta_W g_L g_R (\cos^2\phi_W  k^2-\sin^2 \phi_W v_L^2)}
     {-(g_L^2 - \cos^2 \phi_W \cos^2 \theta_W g_R^2) \cos^2 \phi_W k^2 -
         (g_L^2 - \cos^2\theta_W g_{B-L}^2 \sin^2 \phi_W ) \cos^2\phi_W  v_L^2 +
          \cos^\theta_W 2 g_R^2 v_R^2} \ .
\end{split}\label{eq:ewmix}\end{equation}
where in these expressions, $g_Y$ and $e$ denote the usual hypercharge and electromagnetic coupling constant respectively. Neglecting the $Z$/$Z^\prime$ mixing (\textit{i.e.}, when $\zeta_W \rightarrow 0$), the $Z$ and $Z^\prime$ boson
masses are given through the compact expressions
\begin{equation}
  M_{Z} =  \frac{g_L}{2 \cos\theta_W} \ v
  \qquad{\rm and}\qquad
  M_{Z^\prime} = \frac12 \sqrt{g_{B-L}^2 \sin^2 \phi_W v_L^2 +
     \frac{g_R^2 (\cos^4 \phi_W k^2 + v_R^2)}{\cos^2 \phi_W }} \ .
\label{eq:mz_mzp}
\end{equation}

With a particle spectrum including multiple new charged and neutral scalars and gauge bosons interacting directly with the leptons, the ALRM model can potentially influence many lepton-flavour-violating processes, and impact significantly predictions for the anomalous magnetic dipole moments of the charged leptons. The main purpose of this work is to explore this possibility, and identify regions of the parameter space that are phenomenologically interesting.

\section{The Anomalous Magnetic Moment of the Muon}\label{sec:gm2}
Despite the fact that the gyromagnetic ratio of muon $(g_\mu)$ is precisely 2 at tree level, significant quantum corrections arise from loop contributions. Measurements of the anomalous magnetic moment of the muon, denoted by $a_\mu = \tfrac{g_\mu -2}{2}$, could thus provide an excellent probe of physics Beyond the Standard Model (BSM). In the SM, the predicted value for $a_\mu$, with its uncertainty, has been evaluated by the Muon $g-2$ Theory Initiative in 2020 as~\cite{Czarnecki:2002nt, Melnikov:2003xd, Aoyama:2012wk, Gnendiger:2013pva, Colangelo:2014qya, Kurz:2014wya, Masjuan:2017tvw, Colangelo:2017fiz, Keshavarzi:2018mgv, Colangelo:2018mtw, Hoferichter:2018kwz, Bijnens:2019ghy, Blum:2019ugy, Colangelo:2019uex, Hoferichter:2019mqg, Gerardin:2019vio, Keshavarzi:2019abf, Aoyama:2020ynm}:
\begin{equation}\label{eq:amusm}
    a_\mu^{\text{SM}} =  (116 591 810 \pm 43) \times 10^{-11}\,.
\end{equation}
This value relies on the most up-to-date QED predictions, that are supplemented with electroweak, hadronic vacuum polarisation, and hadronic light-by-light contributions. On the other hand, the latest findings from the E989 experiment at Fermilab \cite{Muong-2:2015xgu} provide the most precise measurement of the anomalous magnetic moment to date, $a_\mu^{\text{expt.}} \text{(FNAL)} =  (116 592 040\pm 54) \times 10^{-11}$~\cite{Muong-2:2021vma}, while the previous result from the Brookhaven National Laboratory is $ a_\mu^{\text{expt.}} \text{(BNL)} =  (116 592 089 \pm 63) \times 10^{-11}$. The combination of these two measurements has yielded a new world average experimental value,  
\begin{equation}
    a_\mu^{\text{expt.}} \text{(Comb)} =  (116592061 \pm 41) \times 10^{-11}\,,
\end{equation}
which confirms and sharpens the discrepancy $\Delta a_\mu$ between experimental observations and the SM prediction~\eqref{eq:amusm} to a level of $4.2 \sigma$. The quantity $\Delta a_\mu$, estimated in~\cite{Muong-2:2021ojo} as
\begin{equation}\label{eq:deltaamu}
    \Delta a_\mu = a_\mu^{\text{expt.}} \text{(Comb)} - a_\mu^{\text{SM}} = (251\pm 59) \times 10^{-11}\,,
\end{equation}
can thus be used as a good motivation for various BSM scenarios. However, recent lattice simulations of the hadronic vacuum polarisation contribution to $a_\mu$ have suggested a reconciliation between theoretical predictions and experimental measurements~\cite{Borsanyi:2020mff, Ce:2022kxy, ExtendedTwistedMass:2022jpw, RBC:2023pvn, FermilabLatticeHPQCD:2023jof}, while also revealing discrepancies in the properties of $e^+e^- \to \pi^+\pi^-$ scattering~\cite{Wittig:2023pcl}. As a result, while the likelihood of large new physics contributions to $a_\mu$ has diminished, unresolved issues remain and continue to motivate BSM explorations in this context. This leaves open the question of whether the deviation~\eqref{eq:deltaamu} should still be interpreted as a sign of new physics affecting hadronic vacuum polarisation. If this interpretation holds, one must keep in mind that such contributions generally worsen electroweak precision fits. Conversely, models predicting smaller contributions to $a_\mu$ may become increasingly attractive in light of the recent lattice findings (and a negative answer to the above question). One-loop electroweak corrections to the muon anomalous magnetic moment were first computed over five decades ago~\cite{Brodsky:1966mv,Burnett:1967zfb,Jackiw:1972jz,Bars:1972pe,Fujikawa:1972fe,Altarelli:1972nc,Bardeen:1972vi}, and comprehensive reviews discussing the computation of both SM and BSM contributions can be found in \cite{Lindner:2016bgg, Aoyama:2020ynm, Yu:2021suw}. Additionally, several studies have explored new physics contributions to the anomalous magnetic moment at the one-loop level within specific theoretical frameworks \cite{Leveille:1977rc, Altmannshofer:2016brv, Gninenko:2001hx, Jana:2020pxx}.

As the ALRM introduces additional (exotic) fermions, scalars and gauge bosons, it is important to  assess their impact on predictions for $a_\mu$, together with the associated consequences on the parameter space of the model. In this work, we analyze one-loop and two-loop (Barr-Zee) contributions in turn.

\subsection{One-loop contributions}
\begin{figure}
    \centering
    \includegraphics[scale=0.36]{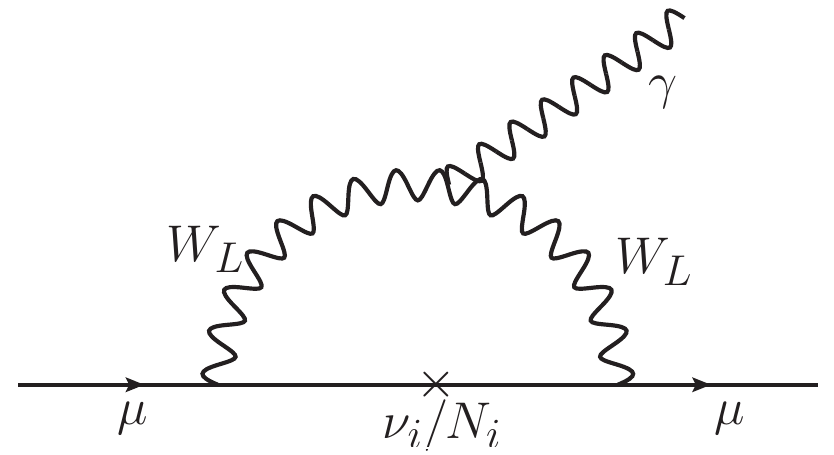} \qquad
    \includegraphics[scale=0.36]{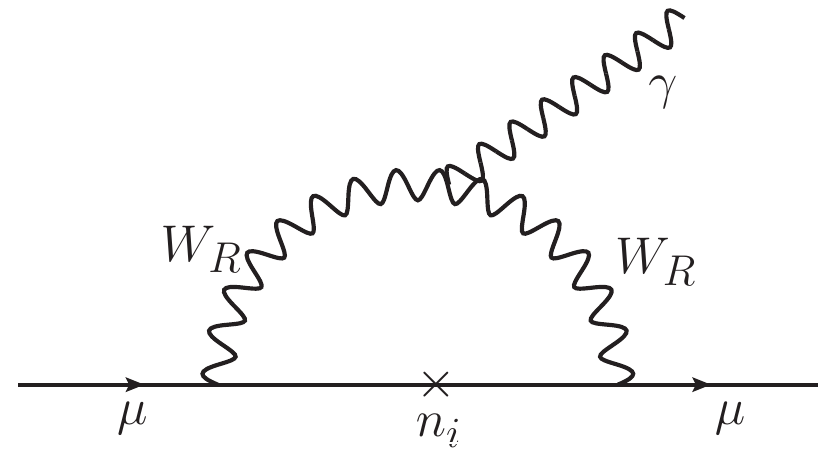} \qquad
    \includegraphics[scale=0.36]{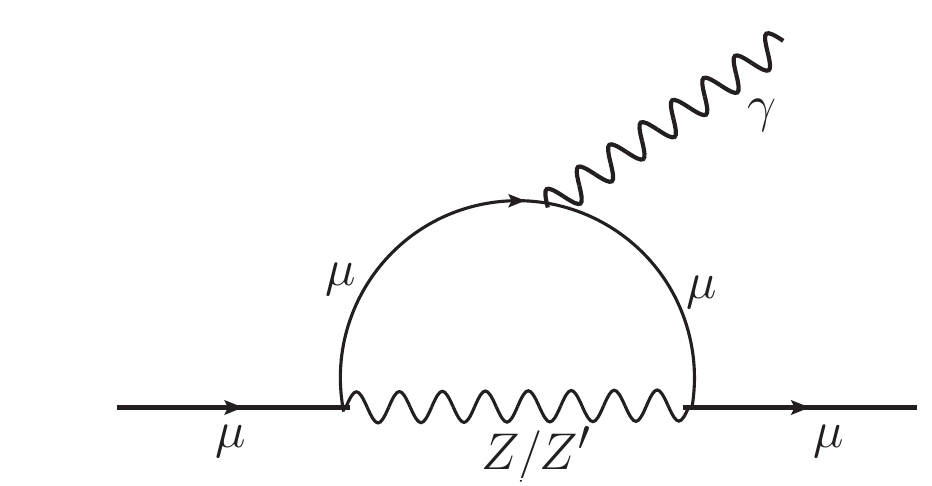}
    \caption{Representative Feynman diagrams contributing to $a_\mu$ and involving gauge couplings of the model's fermions.}
    \label{fig:LFV_Feynman1}
\end{figure}
\begin{figure}
    \centering
    \includegraphics[scale=0.35]{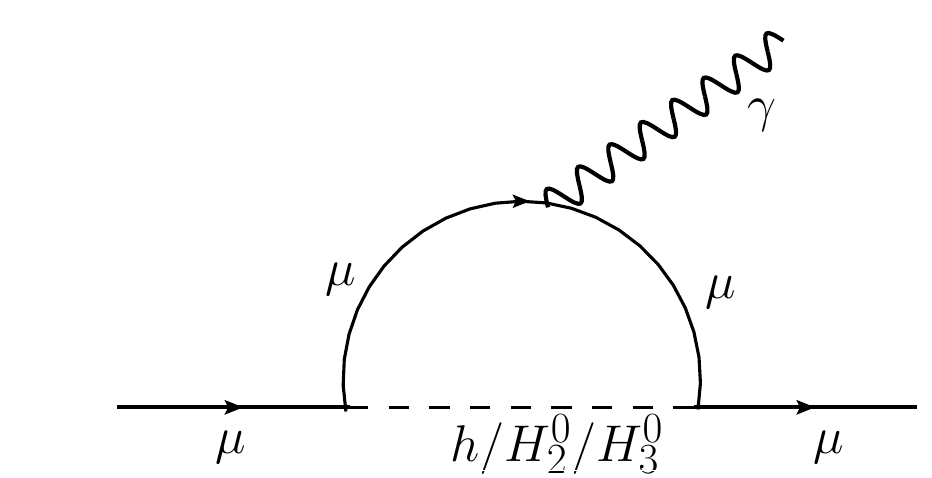} \qquad
    \includegraphics[scale=0.35]{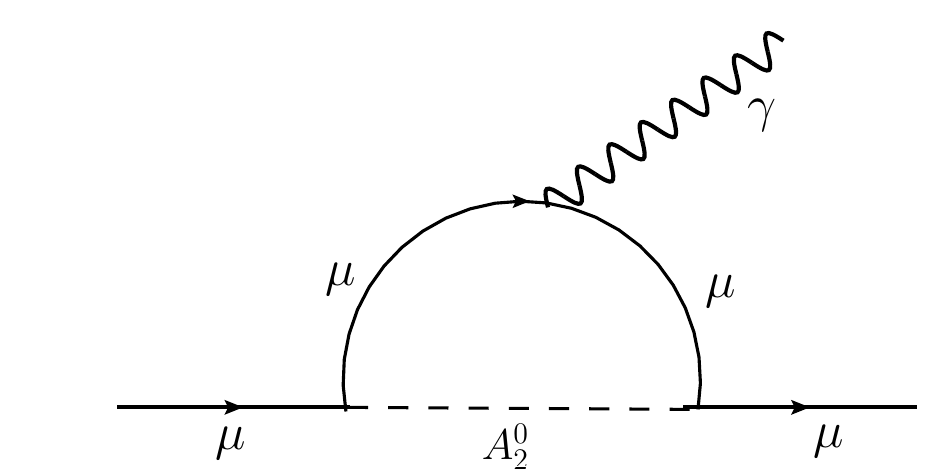}\qquad
    \includegraphics[scale=0.35]{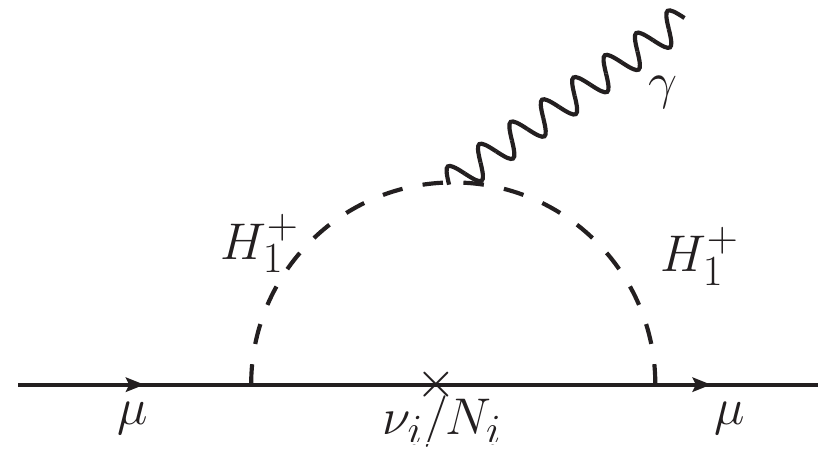} \qquad
    \includegraphics[scale=0.35]{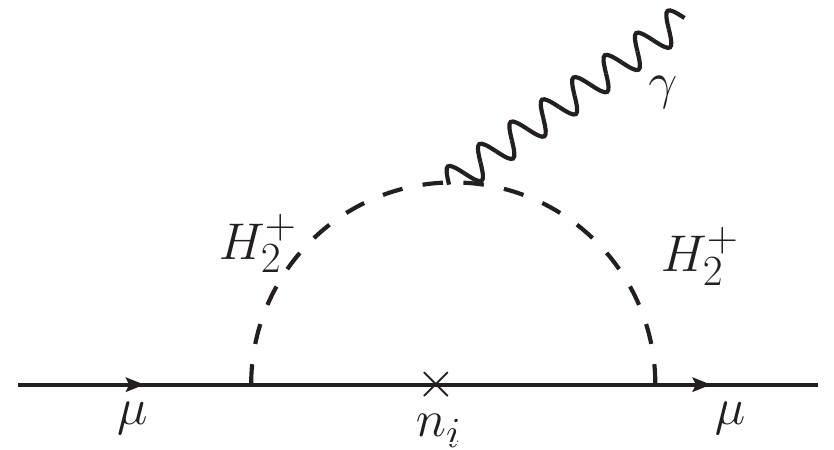}
    \caption{Representative Feynman diagrams contributing to $a_\mu$ and involving the model's scalars.}
    \label{fig:LFV_Feynman2}
\end{figure}

In our model, one-loop contributions involving massive gauge and scalar bosons to $\Delta a_\mu$ arise from the Feynman diagrams given in Figs.~\ref{fig:LFV_Feynman1} and \ref{fig:LFV_Feynman2}, that represent contributions proportional to the model's gauge couplings and those involving the couplings of the model's scalars, respectively. The first set of contributions involving gauge bosons yields~\cite{Lindner:2016bgg,Aoyama:2020ynm,Yu:2021suw}:
\begin{align}
    & \Delta a_\mu (W_L) = \tfrac{1}{4 \pi^2} \left( \tfrac{m^2_\mu}{M^2_{W_L}} \right) \tfrac{5}{6} \tfrac{g_L^2}{4} \left[ \left| \mathcal{V}_{\mu i}^{\nu \nu} \right|^2 + \left| \mathcal{V}_{\mu i}^{\nu N } \right|^2 \right], \nonumber \\
    & \Delta a_\mu (W_R) = \tfrac{1}{4 \pi^2} \left( \tfrac{m^2_\mu}{M^2_{W_R}} \right) \tfrac{5}{6} \tfrac{g_R^2}{4} \left| \mathcal{V}_{\mu i}^{n n } \right|^2 ,\nonumber \\
    & \Delta a_\mu (Z) = - \tfrac{1}{4 \pi^2} \left( \tfrac{m^2_\mu}{M^2_{Z}} \right) \left( \tfrac{e^2}{12} \right) \left[ \tfrac{g_L^2}{g_Y^2} - \tfrac{g_Y^2}{g_L^2} + 4 \right], \nonumber\\
    & \Delta a_\mu (Z^\prime) = - \tfrac{1}{4 \pi^2} \left( \tfrac{m^2_\mu}{M^2_{Z^\prime}} \right) \left( \tfrac{e^2
}{12} \right) \left( \tfrac{g_R^2}{g_Y^2} - \tfrac{g_Y^2}{g_R^2} -1 \right)/\left( \tfrac{1}{g_Y^2} - \tfrac{1}{g_R^2} \right),
\label{eq:amu_oneloop1}\end{align}
while the second set of diagrams involving scalars gives~\cite{Krawczyk:1996sm,Larios:2001ma,Dedes:2001nx,Queiroz:2014zfa,Broggio:2014mna,Lindner:2016bgg,Yu:2021suw}: 
\begin{align}
    & \Delta a_\mu (h) =  \tfrac{1}{4 \pi^2} \left( \tfrac{m^2_\mu}{M^2_{h}} \right) \tfrac{\left|Y^\mu \right|^2}{2} \left| U_{11}^H \right|^2 \left[ - \tfrac{7}{12} - \log \left( \tfrac{m_\mu}{M_{h}} \right) \right], \nonumber\\
    & \Delta a_\mu (H_2^0) =  \tfrac{1}{4 \pi^2} \left( \tfrac{m^2_\mu}{M^2_{H_2^0}} \right) \tfrac{\left|Y^\mu \right|^2}{2} \left| U_{12}^H \right|^2 \left[ - \tfrac{7}{12} - \log \left( \tfrac{m_\mu}{M_{H_2^0}} \right) \right], \nonumber\\
    & \Delta a_\mu (H_3^0) =  \tfrac{1}{4 \pi^2} \left( \tfrac{m^2_\mu}{M^2_{H_3^0}} \right) \tfrac{\left|Y^\mu \right|^2}{2} \left| U_{13}^H \right|^2 \left[ - \tfrac{7}{12} - \log \left( \tfrac{m_\mu}{M_{H_3^0}} \right) \right], \nonumber\\
    & \Delta a_\mu (A_2^0) =  \tfrac{1}{4 \pi^2} \left( \tfrac{m^2_\mu}{M^2_{A_2^0}} \right) \tfrac{\left|Y^\mu \right|^2}{2} \left| U_{11}^A \right|^2 \left[ \tfrac{11}{12} + \log \left( \tfrac{m_\mu}{M_{A_2^0}} \right) \right],\nonumber \\
    & \Delta a_\mu (H_1^+) = - \tfrac{1}{4 \pi^2} \left( \tfrac{m^2_\mu}{M^2_{H_1^+}} \right) \left( \tfrac{1}{48} \right) \left( \tfrac{1}{k^2 + v_L^2} \right) \left[ \left| Y^\mu \right|^2  v_L^2 \left( \left| \mathcal{V}^{\nu \nu}_{\mu i} \right|^2  + \left| \mathcal{V}^{\nu N }_{\mu i} \right|^2 \right) + \left| Y^{\mu }_L \right|^2 k^2 \left( \left| \mathcal{V}^{N \nu}_{\mu i} \right|^2  + \left| \mathcal{V}^{N N}_{\mu i} \right|^2 \right)  \right], \nonumber\\
    & \Delta a_\mu (H_2^+) = - \tfrac{1}{4 \pi^2} \left( \tfrac{m^2_\mu}{M^2_{H_2^+}} \right) \left( \tfrac{1}{48} \right) \left( \tfrac{1}{k^2 + v_R^2} \right) \left| \mathcal{V}^{n n}_{\mu i} \right|^2 \left[ \left| Y^{\mu} \right|^2 v_R^2 + \left| Y^{\mu}_R \right|^2 k^2 \right].
\label{eq:amu_oneloop2}\end{align}

\begin{figure}
    \centering
    \includegraphics[width=0.75\textwidth]{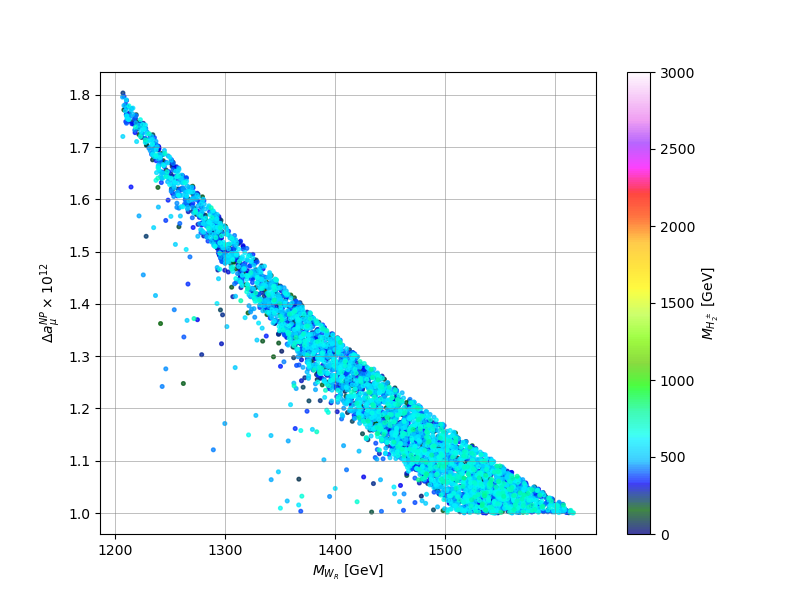}
    \caption{New physics one-loop contribution $\Delta a_\mu^\mathrm{NP}$ to the anomalous magnetic moment of the muon in the ALRM. Each point represents one benchmark explored in our random parameter space scan, and predictions for $\Delta a_\mu^\mathrm{NP}$ are displayed as a function of the mass of the $SU(2)_R$ boson, $M_{W_R}$. In addition, the colour code refers to the mass of the second charged Higgs boson, $M_{H_2}^{\pm}$, and we have only considered scenarios in which the $Z^\prime$ mass satisfies $M_{Z^\prime} > $ 4.5~TeV. For clarity, scenarios for which $\Delta a_{\mu}^\mathrm{NP} < 10^{-12}$ have been omitted. \label{fig:deltaA_MWR} }
\end{figure}

The interaction vertices and the corresponding couplings are listed in the tables given in Appendix~\ref{sec:appendix}, where, as in the rest of this work, we conveniently denote all scotino mass eigenstates by $n\equiv \hat{n}$ to simplify the notation. In order to assess the impact of these diagrams quantitatively, we implement a random scan of the model's parameter space in which we allow the free parameters of the model to vary in the following ranges:
\begin{align}
   {\rm gauge ~~couplings:} &~~~g_L = 0.65, ~~~~g_R \in [0.37,1], \nonumber \\
 {\rm  VEVs:} &~~~ k = 246 ~\text{GeV},~~ v_R \in [6.5, 50] ~\text{TeV}, ~~\tan \beta \in [25, 50], ~~\nonumber \\
   {\rm Trilinear ~couplings:
   } 
   &~~~ {\kappa} \in [-3000,-10] ~\text{GeV}, \nonumber \\
      {\rm Exotic~lepton~masses:} &~~~ m_{n_1, n_2, n_3} \in [1, 10^{5}] ~\text{GeV},  \nonumber\\
       {\rm Neutrino~sector~masses:} & ~~~M_{N_1} = 10^{3} ~\text{GeV}, M_{N_2} = 10^{3.5} ~\text{GeV}, M_{N_3} = 10^{4} ~\text{GeV}, \nonumber \\
       &~~~ m^{\text{lightest}}_\nu \in [10^{-6},0.01] ~\text{eV}, \nonumber \\
 {\rm Other ~Yukawa~couplings}:&~~~ Y^\mu = \tfrac{\sqrt{2}m_\mu}{k} = 6.08 \times 10^{-4}, Y^e = \tfrac{\sqrt{2}m_e}{k} = 2.94 \times 10^{-6},
\label{eq:input1}\end{align} 
where $Y^e\equiv(Y^\ell)_{11}$ and $Y^\mu\equiv(Y^\ell)_{22}$. Furthermore, the active neutrino oscillation parameters are taken such that they lie within $3\sigma$ of their experimental value~\cite{Gonzalez-Garcia:2021dve},
\begin{align}
       &~~~\Delta m^2_{\rm {21}}  \in [6.81, 8.03 ]\times 10^{-5} \mbox{eV}^2,~
       |\Delta m^2_{\rm {31}}| \in  [2.428, 2.597]\times  10^{-3} \mbox{eV}^2,~~\delta_{\rm{CP}} \in [108^\circ, 404^\circ],       \nonumber \\
       &~~~\sin^2\theta_{12} \in [ 0.275, 0.344] ,~\sin^2\theta_{23} \in [ 0.407, 0.620 ],~\sin^2\theta_{13} \in [0.02029, 0.02319]\,,
\label{eq:inputNu}\end{align}
while the heavy neutrino and scotino mixing angles are varied over the range $[0, \pi/2]$ with a zero CP-violating Dirac phase. Considering a normal hierarchical active neutrino mass ordering, the masses of the second and third neutrinos are given by
 \begin{center}
     $m_{\nu_2} = \sqrt{(m_{\nu}^{\rm{lightest}})^2 + \Delta m_{21}^2}$, ~~~  $m_{\nu_3} = \sqrt{(m_{\nu}^{\rm{lightest}})^2 + |\Delta m_{31}|^2}$,
 \end{center}
while the general $3\times 3$ active neutrino mass matrix $m_\nu$ can be determined from
\begin{equation}
    m_\nu = \mathcal{V}^{\nu \nu} m_\nu^{\rm{diag}} (\mathcal{V}^{\nu \nu})^ T
    \qquad\text{with}\qquad 
    m_\nu^{\rm{diag}} = \rm{diag}~[m_{\nu}^{\rm{lightest}}, m_{\nu_2}, m_{\nu_3}].
\end{equation}
The remaining Yukawa couplings, that we consider diagonal, are fully determined once all the parameters above are fixed,
\begin{equation}\begin{split}
    (Y^\ell_L)^{\rm{diag}} = \frac{\sqrt{2 m_\nu^{\rm{diag}} M_N^{\rm{diag}}}}{v_L} \qquad&\text{with}\qquad  M_N^{\rm{diag}} = \rm{diag}~ [M_{N_1}, M_{N_2}, M_{N_3}],\\[.2cm]
  (Y^\ell_R)^{\rm{diag}} = (\mathcal{V}^{nn})^\dagger Y^\ell_R \mathcal{V}^{nn} = \frac{\sqrt{2} m_n^{\rm{diag}}}{v_R} \qquad&\text{with}\qquad  m_n^{\rm{diag}} = \rm{diag}~ [m_{n_1}, m_{n_2}, m_{n_3}].
\end{split}\end{equation}
Correspondingly, the general 3$\times$3 scotino mass matrix can be rewritten as $ m_n = \mathcal{V}^{nn} m_n^{\rm{diag}}  (\mathcal{V}^{nn})^\dagger$. Here scotinos are Dirac fermions such that, without loss of generality, we have assumed that the same unitary mixing matrices could be used to diagonalise its left-handed and right-handed sectors for simplicity. 

In Fig.~\ref{fig:deltaA_MWR}, we present the total contribution of $\Delta a_\mu$ from all diagrams mentioned above against $W_R$-mass, $M_{W_R}$ and after imposing existing LHC bounds on $M_{W_R}$ via existing $Z^\prime$ mass limits (through the associated tree-level mass relations). Consequently, $M_{W_R}$ has to be larger than approximately 1200~GeV~\cite{Frank:2019nid}. In addition, our colour scheme indicates the value of the mass of the second charged Higgs state for each point in the scan. It is found to span a large range of values, although our results show in particular that the lightest charged Higgs $H_2^\pm$ can be light with mass of a few hundreds GeV. On the other hand, all other non-SM neutral Higgs bosons $H_1^0,\, H_2^0$ and $H_3^0$ are found much heavier throughout the whole parameter scan, with masses greater than 5 TeV, and for most of the points the heaviest charged Higgs boson has a mass lying below 1~TeV \cite{Frank:2019nid, Frank:2021ekj}. We opted to focus on the two bosons $W_R$ and $H_2^\pm$ as they are the most relevant ones in light of the analytical results presented in Eqs.~\eqref{eq:amu_oneloop1} and \eqref{eq:amu_oneloop2} and the chosen set of input parameters. Notice that the other charged Higgs, 
$H_1^+$ is much heavier than $H_2^+$.
Our predictions show that ALRM one-loop contributions to $a_\mu$ increase with decreasing $M_{W_R}$ mass values. This different behaviour is inherent to the dependence of the scalar contributions on the various Higgs VEVs, that is absent for diagrams with only gauge-boson exchanges. Nevertheless, for the scanned values of $M_{W_R}$ and $v_R$ consistent with the constraints imposed and for the whole range of variations in the input parameters, the new physics contributions to the anomalous magnetic moment of the muon are still three orders of magnitude smaller than what would be required to get consistency between the experiment and the prediction of the Muon $g-2$ Theory initiative given in Eq.~\eqref{eq:amu_oneloop1}. On the other hand, accounting for the newest lattice results would not challenge the viability of the model, despite a rich leptonic sector.

While these findings are encouraging, we now proceed to investigate ALRM two-loop contributions to $a^\mu$, as the latter could be significant for new physics setups allowing for new heavy fermions to circulate in the loops.

\subsection{Two-loop contributions}

\begin{figure}
    \centering
    \includegraphics[scale=0.45]{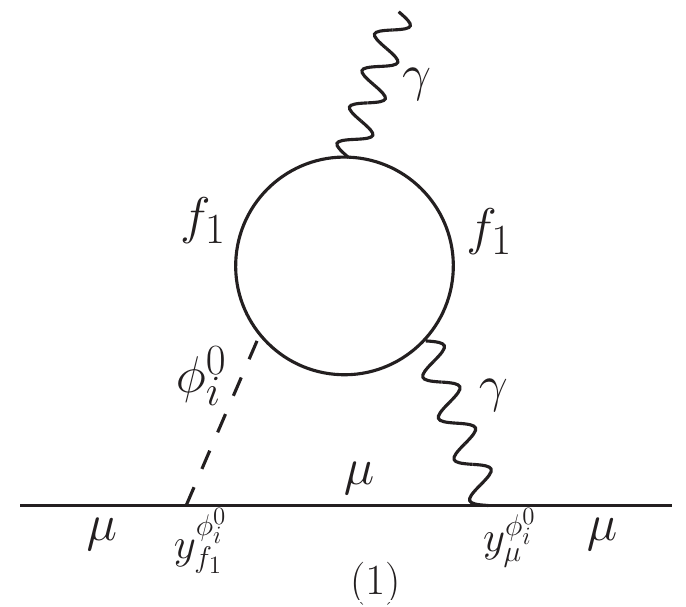} \qquad
    \includegraphics[scale=0.45]{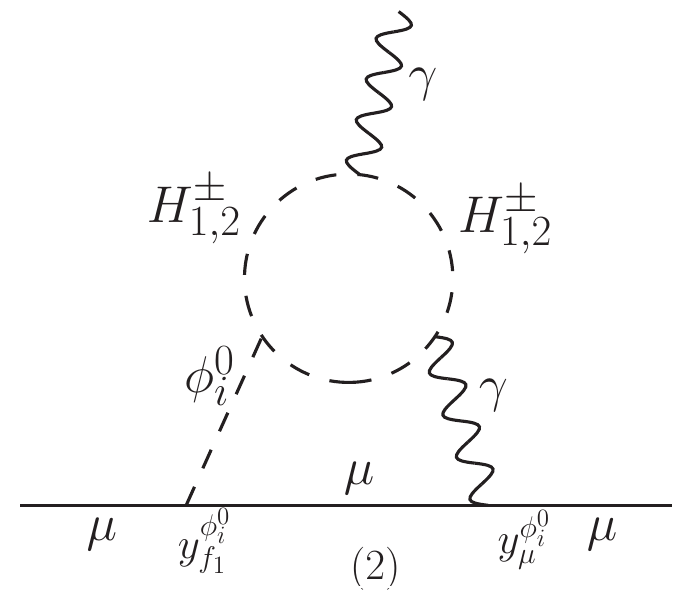} \qquad
    \includegraphics[scale=0.45]{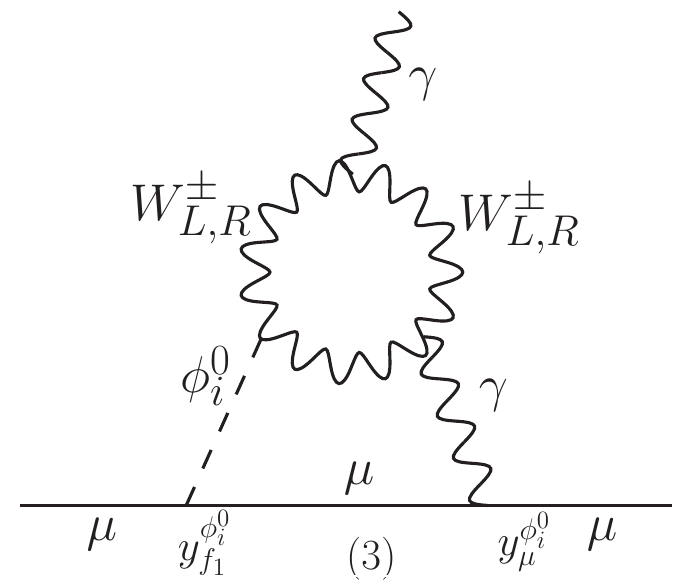} \qquad
    \includegraphics[scale=0.45]{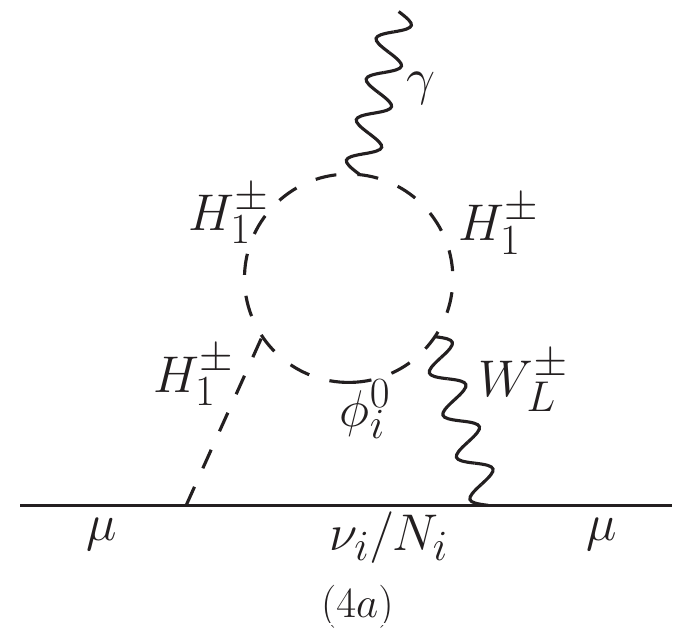}\qquad
    \includegraphics[scale=0.45]{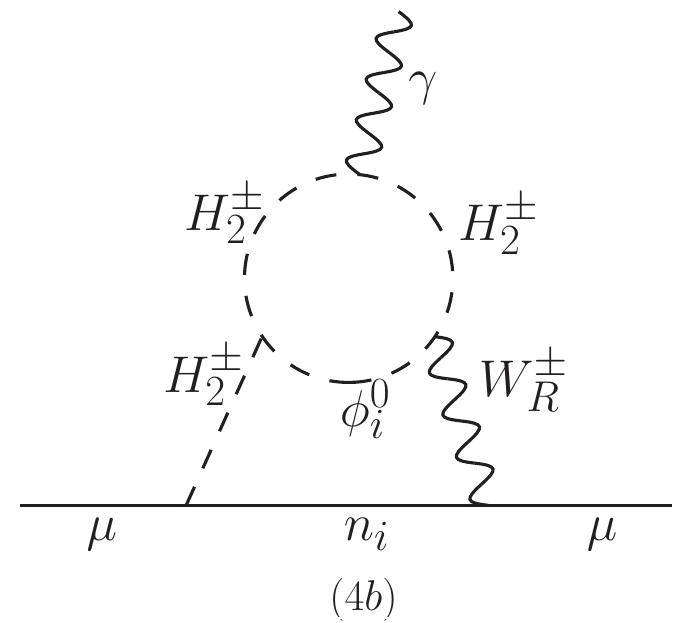}\qquad
    \includegraphics[scale=0.45]{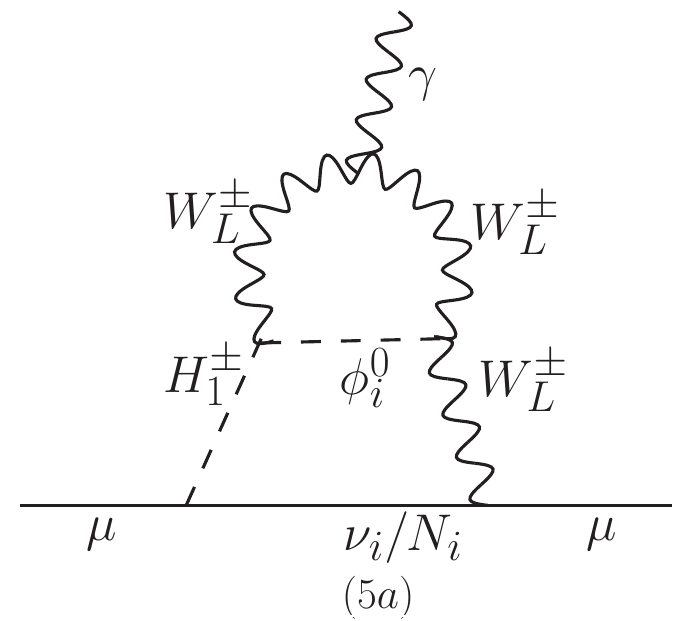}\qquad
    \includegraphics[scale=0.45]{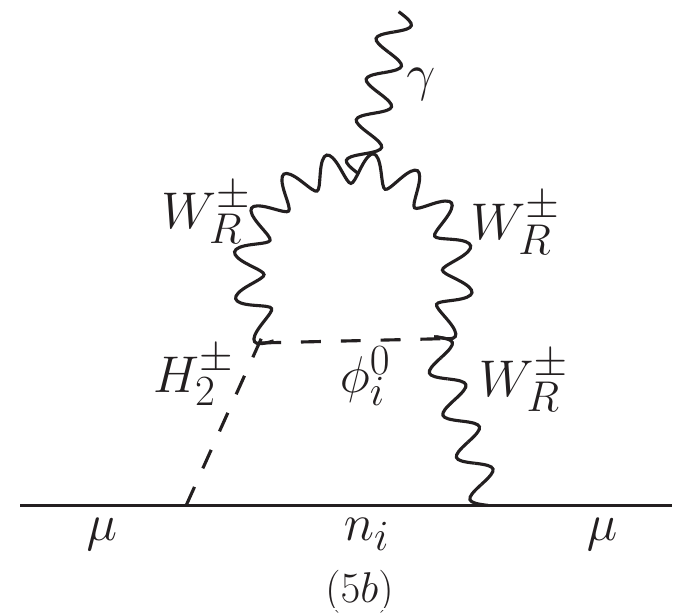} \qquad
    \includegraphics[scale=0.45]{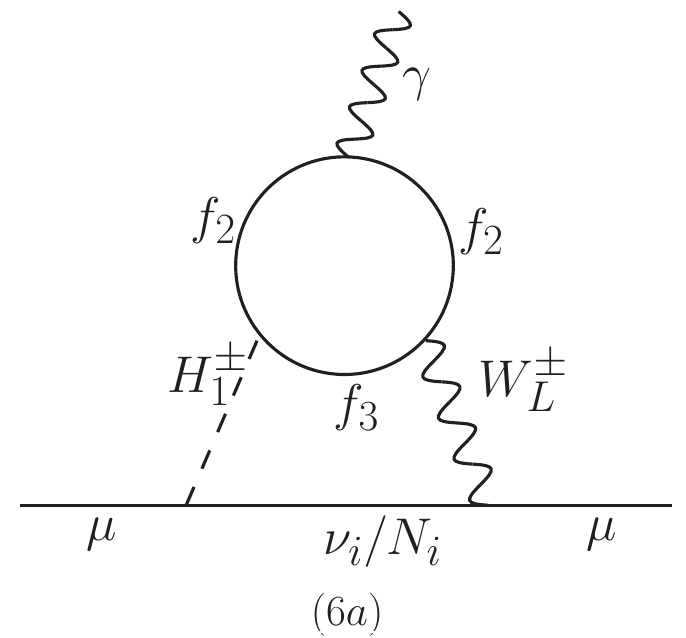} \qquad
    \includegraphics[scale=0.45]{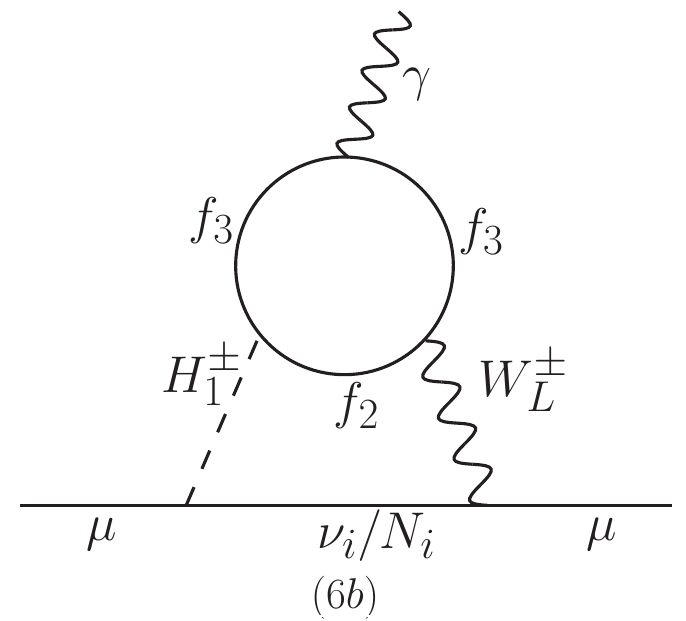}
    \caption{Representative new physics Feynman diagrams contributing to the anomalous magnetic moment of the muon at two loop in the ALRM.
    \label{fig:2loop1}}
\end{figure}

The different class of (Barr-Zee) \cite{Bjorken:1977vt, Barr:1990vd} Feynman diagrams responsible for two-loop ALRM contributions to the muon anomalous magnetic moment are shown in Fig.~\ref{fig:2loop1}. Analytical expressions for the first three class of diagrams, $\Delta a_\mu^{(i)}$ with $i=1, 2, 3$, are given by \cite{Chang:2000ii,Dedes:2001nx,Larios:2001ma,Gunion:2008dg,Broggio:2014mna,Wang:2014sda,Wang:2014sda,Ilisie:2015tra}
\begin{equation}\begin{split}
  \Delta a^{(1)}_\mu =&\ \tfrac{e^2 m_\mu^2}{128 \pi^4 v^2} \sum_{\phi_i^0,f_1}  N_C^{f} ~Q_{f}^2 ~Y_{f}^{\phi^0_i}  Y_\mu^{\phi^0_i} ~G_{\phi^0_i}^{(1)} \left( \tfrac{m_{f_1}^2}{M_{\phi^0_i}^2}\right), \\
  \Delta a^{(2)}_\mu =&\ \tfrac{e^2 m_\mu^2}{64 \pi^4 M_{\phi^0_i}^2} \sum_{\phi_i^0} ~Y_\mu^{\phi^0_i} ~\lambda_{\phi^0_i H^+_{1,2} H^-_{1,2}} ~G^{(2)} \left( \tfrac{M_{H^\pm_{1,2}}^2}{M_{\phi^0_i}^2}\right),\\
  \Delta a^{(3)}_\mu =&\ \tfrac{e^2 m_\mu^2}{128 \pi^4 v^2 } \sum_{\phi_i^0}~ Y_\mu^{\phi^0_i} Y_W^{\phi^0_i} G^{(3)} \left( \tfrac{M_{W_{L,R}}^2}{M_{\phi^0_i}^2}\right).
\end{split}\end{equation}
Here, the $M_i$ parameters stand for the physical masses of the different bosonic states, the $m_i$ parameters for the masses of the quarks and charged leptons in the model, and the Yukawa couplings $Y_{f}^{\phi^0_i} $ and  $Y_\mu^{\phi^0_i} $ are the strengths of the $\bar{f} f \phi_i^0$ and $\overline{\mu} \mu \phi_i^0$ interactions, respectively, with the generic fermion $f \equiv t,b, \tau, d^\prime, s^\prime, b^\prime$ and the generic scalar state $\phi_i^0 \equiv h, H_2^0, H_3^0, A_2^0$. Explicit expressions for these couplings can be found in the  tables given in Appendix~\ref{sec:appendix}. In addition, these two-loop contributions depend on the number of colours $N_C^{f}$ related to a specific fermion $f$, the related electric charge $Q_{f}$ and the three kinds of $G$ functions defined by
\begin{equation}\begin{split}
  G_{h/ H_2^0/ H_3^0}^{(1)} (r) =&\ r \int_0^1 dx \tfrac{2x(1-x)-1}{x(1-x)-r} \text{ln} \left[\tfrac{x(1-x)}{r} \right], \\
  G_{A_2^0}^{(1)} (r) =&\ r \int_0^1 dx \tfrac{1}{x(1-x)-r} \text{ln} \left[\tfrac{x(1-x)}{r} \right] \\
  G^{(2)} (r) =&\ \int_0^1 dx \tfrac{x(x-1)}{x(1-x)-r} \text{ln} \left[\tfrac{x(1-x)}{r} \right] \\ 
  G^{(3)} (r) =&\ \tfrac{1}{2} \int_0^1 dx \tfrac{x[3x(4x-1)+10]r-x(1-x)}{x(1-x)-r} \text{ln} \left[\tfrac{x(1-x)}{r} \right]\,.
\end{split}\end{equation}
The contributions of the diagram of category $4a$ and $4b$ (first two diagrams of the second line in Fig.~\ref{fig:2loop1}) read \cite{Ilisie:2015tra}
\begin{equation}\begin{split}
     \Delta a^{(4a)}_\mu = &\ \tfrac{ m_\mu^2}{256 \pi^4} \tfrac{1}{\left( M^2_{H^\pm_1} - M_{W_L}^2 \right)} \sum_{\phi_i^0}~( Y_{\mu \nu_i H_1^\pm} g_{\mu \nu_i W_L^\pm} + Y_{\mu N_i H_1^\pm} g_{\mu N_i W_L^\pm}) \lambda_{\phi^0_i H^+_{1} H^-_{1}} g_{W_L^- H_1^+ \phi^0_i}  \\
    &\quad \times \int_0^1 dx~ x^2 (x-1)  \left[ F \left(1, \tfrac{M_{\phi^0_i}^2}{M^2_{H^\pm_1}} \right) - F \left( \tfrac{M^2_{H^\pm_1}}{M^2_{W_L}}, \tfrac{M_{\phi^0_i}^2}{M^2_{W_L}} \right) \right], \\
    \Delta a^{(4b)}_\mu = & \tfrac{ m_\mu^2}{256 \pi^4} \tfrac{1}{\left( M^2_{H^\pm_2} - M_{W_R}^2 \right)} \sum_{\phi_i^0} ~ Y_{\mu n_i H_2^\pm} \lambda_{\phi^0_i H^+_{2} H^-_{2}} g_{W_R^- H_2^+ \phi^0_i} g_{\mu n_i W_R^\pm} \\
    &\quad \times \int_0^1 dx~ x^2 (x-1)  \left[ F \left(1, \tfrac{M_{\phi^0_i}^2}{M^2_{H^\pm_2}} \right) - F \left( \tfrac{M^2_{H^\pm_2}}{M^2_{W_R}}, \tfrac{M_{\phi^0_i}^2}{M^2_{W_R}} \right) \right],
\end{split}\end{equation}
and depend on some multiscalar and gauge couplings given explicitly in Appendix~\ref{sec:appendix}, as well as on the function $F$ defined by 
\begin{equation}
  F(r_1, r_2) = \int_0^1 dx~\tfrac{\text{ln} \left( \tfrac{r_1 x + r_2 (1-x)}{x(1-x)} \right)}{x(1-x) - r_1 x - r_2 (1-x)}\,.
\end{equation}
The same function also appears in the contributions from the last four diagrams, labeled $5a$, $5b$, $6a$ and $6b$, that are given by \cite{Ilisie:2015tra}
\begin{equation}\begin{split}
     &\Delta a^{(5a)}_\mu = \tfrac{ m_\mu^2}{256 \pi^4 v^2} \tfrac{1}{\left( M^2_{H^\pm_1} - M_{W_L}^2 \right)} \sum_{\phi_i^0}~( Y_{\mu \nu_i H_1^\pm} g_{\mu \nu_i W_L^\pm} + Y_{\mu N_i H_1^\pm} g_{\mu N_i W_L^\pm}) g_{W_L^- H_1^+ \phi^0_i} g_{W_L^+ W_L^- \phi^0_i} \\
    &\quad \times \int_0^1 dx~ x^2  \left[ \left( M_{H_{1}^\pm}^2 \!+\! M_{W_L}^2 \!-\! M_{\phi_i^0}^2 \right) (1\!-\!x) \!-\! 4 M_{W_L}^2 \right] \left[ F \left( \tfrac{M^2_{W_L}}{M^2_{H^\pm_1}}, \tfrac{M_{\phi^0_i}^2}{M^2_{H^\pm_1}} \right) - F \left(1, \tfrac{M_{\phi^0_i}^2}{M^2_{W_L}} \right)  \right], \\
    &\Delta a^{(5b)}_\mu = \tfrac{ m_\mu^2}{256 \pi^4 v^2} \tfrac{1}{\left( M^2_{H^\pm_2} - M_{W_R}^2 \right)} \sum_{\phi_i^0}~ Y_{\mu n_i H_2^\pm} g_{\mu n_i W_R^\pm} g_{W_R^- H_2^+ \phi^0_i} g_{W_R^+ W_R^- \phi^0_i} \\
    &\quad \times \int_0^1 dx~ x^2  \left[ \left( M_{H_2^\pm}^2 \!+\! M_{W_R}^2 \!-\! M_{\phi_i^0}^2 \right) (1\!-\!x) \!-\! 4 M_{W_R}^2 \right] \left[ F \left( \tfrac{M^2_{W_R}}{M^2_{H^\pm_2}}, \tfrac{M_{\phi^0_i}^2}{M^2_{H^\pm_2}} \right) \!-\! F \left(1, \tfrac{M_{\phi^0_i}^2}{M^2_{W_R}} \right)  \right] \,,\\
    &\Delta a^{(6a)}_\mu = \tfrac{ m_\mu^2}{256 \pi^4 v^2} \tfrac{N_C}{\left( M^2_{H^\pm_1} - M_{W_L}^2 \right)} \int_0^1 dx ( Y_{\mu \nu_i H_1^\pm} g_{\mu \nu_i W_L^\pm} + Y_{\mu N_i H_1^\pm} g_{\mu N_i W_L^\pm})g_{tbW_L^\pm}~ g_{tbH_1^\pm} \\
    &\quad \times  \left[ Q_t x \!+\! Q_b (1\!-\!x) \right]  \Biggl[ m_b^2 x(1\!-\!x) \!+\! m_t^2 x(1\!+\!x) \Biggr] \left[F \left( \tfrac{m^2_{t}}{M^2_{H_1^\pm}}, \tfrac{m_{b}^2}{M^2_{H_1^\pm}} \right) \!-\! F \left( \tfrac{m_{t}^2}{M^2_{W_L}}, \tfrac{m_{b}^2}{M^2_{W_L}} \right)  \right], \\
    &\Delta a^{(6b)}_\mu = \tfrac{ m_\mu^2}{256 \pi^4 v^2} \tfrac{N_C}{\left( M^2_{H^\pm_2} - M_{W_R}^2 \right)} \sum_{q'=d', s', b'} \int_0^1 dx Y_{\mu n_i H_2^\pm} g_{\mu n_i W_R^\pm} g_{tq^\prime W_R^\pm}~ g_{tq^\prime H_2^\pm} \\
    &\quad \times  \left[ Q_{t} x \!+\! Q_{q^\prime} (1\!-\!x) \right]  \Biggl[ m_t^2 x(1\!+\!x) \!+\! m_{q^\prime}^2 x(1\!-\!x) \Biggr] \ \left[F \left( \tfrac{m^2_t}{M^2_{H_2^\pm}}, \tfrac{m_{q^\prime}^2}{M^2_{H_2^\pm}} \right) \!-\! F \left( \tfrac{m_t^2}{M^2_{W_R}}, \tfrac{m_{q^\prime}^2}{M^2_{W_R}} \right)  \right]\,.
\end{split}
\end{equation}

\begin{figure}
    \centering 
    \includegraphics[width=.75\textwidth]{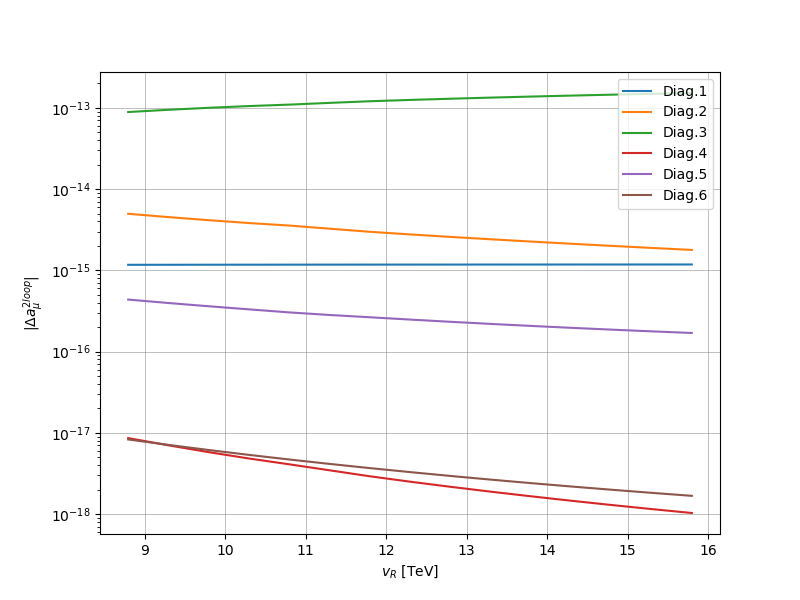}
    \caption{Two-loop contributions to the anomalous magnetic moment of the muon as a function of $v_R$, after separating the impact of the different set of diagrams shown in Fig.~\ref{fig:2loop1} and for the benchmark scenario defined in Eq.~\eqref{eq:bench2loop}.\label{fig:Deltaa_vR_2loop}}
\end{figure}

We now have all the ingredients to evaluate the total contribution to the anomalous magnetic dipole moment of the muon, that simply consists in the sum of the contributions from all one-loop and two-loop individual diagrams listed above,
\begin{equation}
    \Delta a_\mu^{\text{total}} = \Delta a^{\text{1-loop}}_\mu + \Delta a^{\text{2-loop}}_\mu .
\end{equation}
In Fig.~\ref{fig:Deltaa_vR_2loop}, we display the two-loop contributions from each subset of diagrams of Fig~\ref{fig:2loop1}, as a function of $v_R$ while keeping other parameters fixed to the following values:
\begin{equation}\label{eq:bench2loop}\begin{split}
   &\tan\beta = 30\,,\qquad
   g_R = 0.386\,,\\
   &m_{d^{'}} = 100~\mathrm{GeV}\,,\qquad
   m_{s^{'}} = 500~\mathrm{GeV}\,,\qquad
   m_{b^{'}} = 150~\mathrm{GeV}\,,\\
   &m_{n_1} = 1412~\mathrm{GeV}\,,\qquad
   m_{n_2} = 1290~\mathrm{GeV}\,,\qquad
   m_{n_3} = 174~\mathrm{GeV}\,,\qquad
   \kappa = -87.6~\mathrm{GeV}\,,\\
   &\theta'_{12} = \theta'_{13} = 4.7^{\circ}\,,\qquad
   \theta'_{23} = 9.5^{\circ}\,.
\end{split}\end{equation}
This allows to illustrate, for a given choice of free parameters (other choices lead to similar conclusions), that despite of the presence of several new particles the ALRM does not predict significant contribution to $\Delta a_\mu$, neither at one loop nor at two loops.
 
\section{Lepton flavour Violating Processes}
\label{sec:LFV}
In this section, we present predictions for several LFV processes within the ALRM framework. We choose processes that are scrutinised by current high-energy physics experiments so that associated measurements could serve as indirect probes of the model's parameter space. We specifically focus on muon radiative decays and $\mu-e$ conversion processes in Sections~\ref{sec:muegam} and \ref{sec:muconv}, respectively, as these are expected to provide the most stringent constraints. Nonetheless, the calculations presented below can be straightforwardly applied to rare tau decays like $\tau \to \mu \gamma$ or $\tau \to e \gamma$. 

\begin{table}
  \renewcommand{\arraystretch}{1.2}\setlength\tabcolsep{8pt}
  \centering
  \begin{tabular}{c | cc}
    Process & Present bound & Future sensitivity \\
    \hline
    BR $(\mu \to e \gamma)$ & $3.1 \times 10^{-13}$ \cite{Chiappini:2023luv} & $6 \times 10^{-14}$ \cite{Baldini:2013ke} \\
    \hline 
    CR $(\mu \rm{Au} \to e \rm{Au})$ & $7 \times 10^{-13}$ \cite{SINDRUMII:2006dvw} & - \\
    CR $(\mu \rm{Ti} \to e \rm{Ti})$ & $4.3 \times 10^{-12}$ \cite{SINDRUMII:1996kid} & - \\
    CR $(\mu \rm{Al} \to e \rm{Al})$ & - & $10^{-15} - 10^{-17}$ \cite{Pezzullo:2017iqq} \\
  \end{tabular}
  \caption{Current experimental bounds at 90\% confidence level on the branching ratio (BR) associated with the muon decay $\mu \to e \gamma$ and on the $\mu - e$ conversion rates (CR) in several nuclei, shown together with expected future sensitivities.\label{tab:LFV}}
\end{table}

In many BSM scenarios that account for neutrino masses, the GIM suppression inherent to the SM contributions to LFV processes is weakened due to the mixing between left-handed and right-handed neutrino states, leading to significantly enhanced rates~\cite{Ilakovac:1994kj, Deppisch:2004fa, Deppisch:2005zm, Ilakovac:2009jf, Alonso:2012ji, Dinh:2012bp, Ilakovac:2012sh, Abada:2012cq, Lee:2013htl}. In our model, this mixing is forbidden, and the usual BSM enhancement from loops of $W$ bosons and extra neutrinos is absent. However, additional fields still contribute such that potential enhancement may arise from loops involving charged Higgs bosons $H_{1,2}^{\pm}$ and various neutrino states ($\nu$, $N$ and $n$). Therefore, it is valuable to compare the ALRM predictions with current experimental limits and future sensitivities, which we summarize in Table~\ref{tab:LFV}, to get indirect insights on which ALRM scenarios could be phenomenologically viable.

\subsection{Flavour-violating muon decay \texorpdfstring{$\mu \to e \gamma$}{mu to e gamma}}\label{sec:muegam}

Within the SM, $\mu \to e \gamma$ decays are forbidden due to the absence of flavour-changing-neutral currents and the conservation of the lepton flavor. However, in simple BSM extensions featuring neutrino masses and mixings, this process can occur via higher-order loop diagram contributions. In this case, one-loop effects mediated by diagrams comprising neutrinos and $W$ bosons increase the decay rate, although its value is still negligibly small,
\begin{equation}\label{eq:tinybr}
  \mathrm{BR}(\mu\to e\gamma)^{\mathrm{SM}+\nu} \le 10^{-54}\,.
\end{equation}
This is consequently far beyond the reach of any current~\cite{Chiappini:2023luv} or future experiment (like MEG II~\cite{Baldini:2013ke}), as shown in Table~\ref{tab:LFV}, which makes this process essentially undetectable in near future.

\begin{figure}
    \centering
    \includegraphics[scale=0.35]{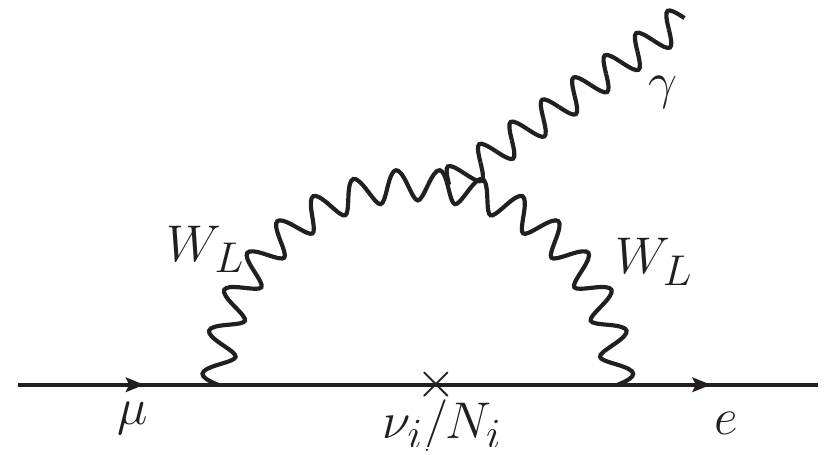} 
    \qquad
    \includegraphics[scale=0.35]{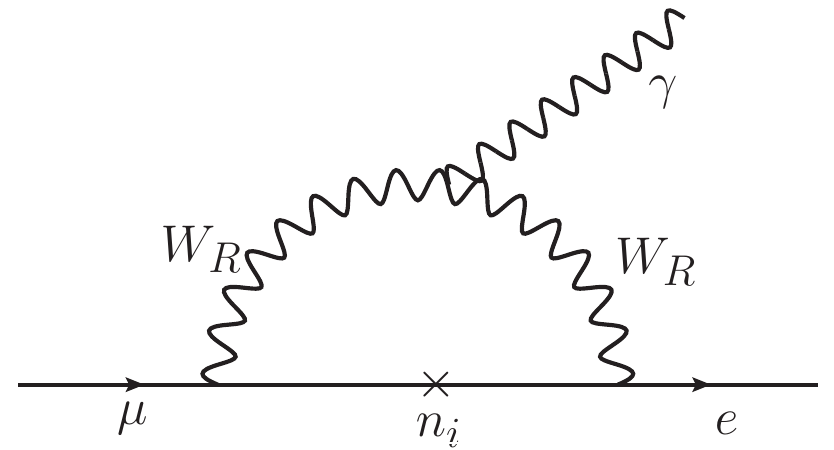}
    \qquad
    \includegraphics[scale=0.35]{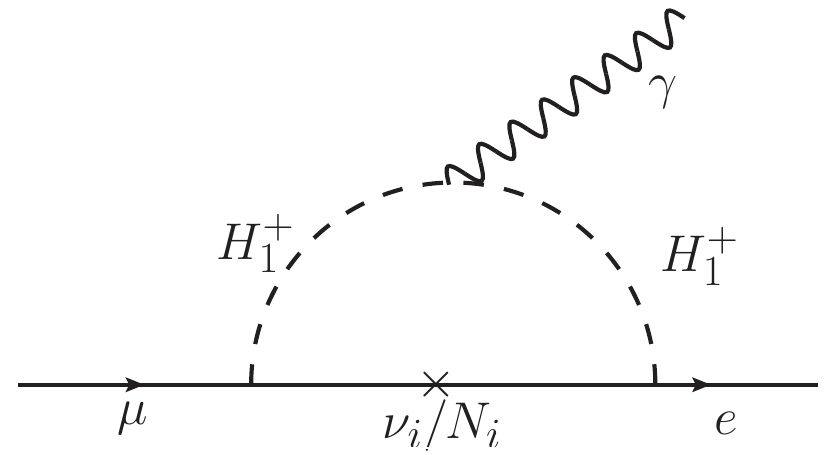}    
    \qquad
    \includegraphics[scale=0.35]{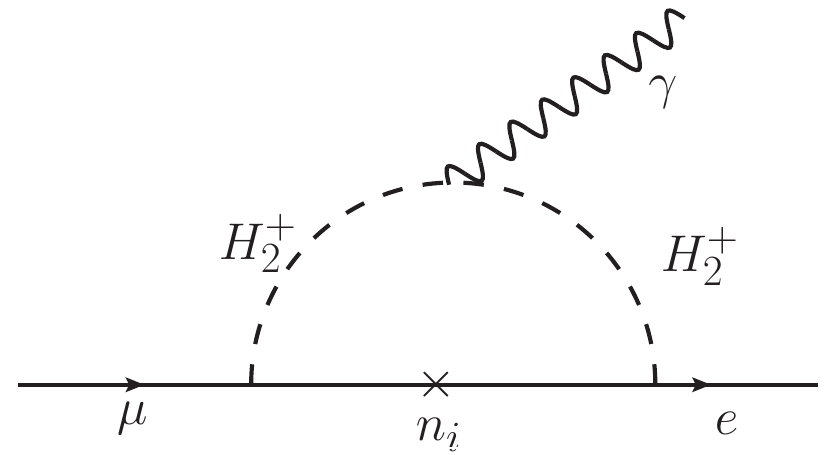}
    \caption{The Feynman diagrams responsible for $\mu \to e \gamma $ in the ALRM model.}
    \label{fig:LFV_muegAll}
\end{figure}
The amplitude relevant for the $\mu \rightarrow e \gamma$ process can be formally written as
\begin{equation}
    \mathcal{A} (\mu \rightarrow e \gamma) = i \epsilon_\nu^\ast(q)~\overline{u}_e (p')\,  \sigma^{\nu \mu}q_\mu \left[ \mathcal{A}_R P_R + \mathcal{A}_L P_L\right]u_\mu (p)\,,
\end{equation}
where $p$, $p'$ and $q$ are the four-momenta of the muon, electron and  photon, respectively. Moreover, following standard notation, $P_{L,R}=\frac{1}{2}\left(1\mp \gamma^5\right)$ represent the left-handed and right-handed chirality projectors, $\sigma^{\mu\nu} = \tfrac{i}{2} [\gamma^\mu, \gamma^\nu]$, while $\bar u_e$, $u_\mu$ stand for the electron and muon spinors and $\epsilon_\nu^\ast$ for the photon polarisation vector. Squaring and averaging this amplitude allows for the computation of the associated partial decay width, which reads, in terms of the  loop factors ${\cal A}_L$ and ${\cal A}_R$,
\begin{equation}
    \Gamma (\mu \rightarrow e \gamma) = \tfrac{m_\mu^3}{16 \pi} \Big( |\mathcal{A}_L|^2 + |\mathcal{A}_R|^2 \Big)\,.
\end{equation}
In this expression, the electron mass $m_e$ has been neglected while $m_\mu$ is the muon mass. In the class of models considered in this work, the presence of a Majorana mass term for the right-handed neutrino state $\nu_R$ leads to additional diagrams mediating the muon decay into an electron-photon final state, via loops comprising neutrinos and $W$ bosons. These diagrams, represented by the first diagram in Fig.~\ref{fig:LFV_muegAll}, are similar to those emerging in the SM$+\nu$ case. The associated contribution is therefore expected to be negligible and no different from Eq.~\eqref{eq:tinybr}, \textit{i.e}\ of ${\cal O}(10^{-54})$. Nevertheless, several other contributions exist in ALRM models, as illustrated by the other Feynman diagrams in Fig.~\ref{fig:LFV_muegAll}. These diagrams open $\mu\to e\gamma$ decay via the exchange of additional (possibly light) charged Higgs bosons $H_{1,2}^\pm$ and charged gauge boson $W_R$. 

The first two diagrams in Fig.~\ref{fig:LFV_muegAll} contribute to the two form factors ${\cal A}_L$ (for diagrams with a $W$-boson exchanges) and ${\cal A}_R$ (for diagrams with a $W_R$-boson exchange), the corresponding amplitude being given by
\begin{equation}\begin{split}
    \mathcal{A}^{W_L}_{\nu_i} =&\ \tfrac{g_L^2}{2} \tfrac{e m_\mu}{64 \pi^2} \tfrac{1}{ M_{W_L}^2} \sum_i \mathcal{V}^{\nu \nu}_{ei} \mathcal{V}^{ \nu \nu \ast}_{\mu i} F_1 \left( \tfrac{m_{\nu_i}^2}{M_{W_L}^2}\right),\\
    \mathcal{A}^{W_L}_{N_i} =&\ \tfrac{g_L^2}{2} \tfrac{e m_\mu}{64 \pi^2} \tfrac{1}{ M_{W_L}^2} \sum_i \mathcal{V}^{\nu N}_{ei} \mathcal{V}^{ \nu N \ast}_{\mu i} F_1 \left( \tfrac{M_{N_i}^2}{M_{W_L}^2}\right),\\
    \mathcal{A}^{W_R}_{n_i} =&\ \tfrac{g_R^2}{2} \tfrac{e m_\mu}{64 \pi^2} \tfrac{1}{ M_{W_R}^2} \sum_i \mathcal{V}^{n n}_{ei} \mathcal{V}^{n n \ast}_{\mu i} F_1 \left( \tfrac{M_{n_i}^2}{M_{W_R}^2}\right),
\end{split}\end{equation}
with the loop function
\begin{equation}
    F_1 (x) = \tfrac{10 -43x+78x^2-49x^3+18x^3 \text{Log}(x)+4x^4}{6(1-x)^4}.
\end{equation}
While ${\cal A}_{\nu_i}^{W_L}$ is negligible owing to both GIM suppression and the smallness of the light-neutrino masses ($m_{\nu}\ll M_{W_L}$), we further assume that light-heavy neutrino mixings $({\cal V}^{\nu N})$ are very small (to accommodate neutrino data) so that ${\cal A}_{N_i}^{W_L}$ is even more insignificant. On the other hand, the scotino mass can be large, which then cancel any suppression in diagrams exhibiting $W_R$ boson exchanges. By virtue of the properties of the function $F_1$, 
\begin{equation}
   F_1(x \rightarrow \infty ) = \tfrac{2}{3} +  \tfrac{3 \text{log}(x)}{x}, ~~~  F_1(x \rightarrow 0 ) = \tfrac{5}{3} - \tfrac{1}{2}x, ~~~ F_1(x \rightarrow 1 ) = \tfrac{17}{12} + \tfrac{3}{20} (1-x),
\end{equation}
the relevant amplitude can be simplified according to the mass hierarchy between the scotino and the $W_R$ boson. We hence have 
\begin{equation}\begin{split}
    m_{n_i} \ll M_{W_R} \quad \to \quad \mathcal{A}_{n_i}^{W_R} =&\ - \tfrac{g_R^2}{4} \tfrac{e m_\mu}{64 \pi^2} \tfrac{1}{M_{W_R}^4} \sum_i m_{n_i}^2 ~\mathcal{V}^{n n}_{ei} \mathcal{V}^{n n \ast}_{\mu i}\\
    = &\ \tfrac{g_R^2}{4} \tfrac{e m_\mu}{64 \pi^2} \tfrac{1}{ M_{W_R}^4} \left(\Delta m_{n12}^2~ \mathcal{V}^{n n}_{e2} \mathcal{V}^{n n \ast}_{\mu 2}+\Delta m_{n13}^2~ \mathcal{V}^{n n}_{e3} \mathcal{V}^{n n \ast}_{\mu 3}\right)\,,\\
    m_{n_i} \gg M_{W_R} \quad \to \quad \mathcal{A}_{n_i}^{W_R} =&\ \tfrac{3 g_R^2}{2} \tfrac{e m_\mu}{64 \pi^2} \sum_i  \tfrac{1}{m_{n_i}^2} \text{Log}\left( \tfrac{m_{n_i}^2}{M_{W_R}^2}\right) ~ \mathcal{V}^{n n}_{ei} \mathcal{V}^{n n \ast}_{\mu i}\,,\\
    m_{n_i} \sim M_{W_R}  \quad \to \quad \mathcal{A}_{n_i}^{W_R} =&\ - \tfrac{3 g_R^2 }{2} \tfrac{e m_\mu}{1280 \pi^2} \tfrac{1}{M_{W_R}^4} \sum_i  { m_{n_i}^2}~ \mathcal{V}^{n n}_{ei} \mathcal{V}^{n n \ast}_{\mu i}\\
    =&\  \tfrac{3g_R^2}{40} \tfrac{e m_\mu}{64 \pi^2} \tfrac{1}{ M_{W_R}^4} \left(\Delta m_{n12}^2~ \mathcal{V}^{n n}_{e2} \mathcal{V}^{n n \ast}_{\mu 2}+\Delta m_{n13}^2~ \mathcal{V}^{n n}_{e3} \mathcal{V}^{n n \ast}_{\mu 3}\right)\,,
\end{split} \label{eq:scotino_amplitudes}\end{equation}
with $\Delta m_{n_{12}}^2=m_{n_1}^2-m_{n_2}^2$ and $\Delta m_{n_{13}}^2=m_{n_1}^2-m_{n_3}^2$. When the particle spectrum is such that $m_{n_i} \gg M_{W_R}$, all contributions are suppressed by factor of  $1/m_{n_i}^2$. This is not the case for other spectrum configurations. In particular, when $m_{n_i} \approx M_{W_R}$, the amplitude could get relatively large, provided that the scotino spectrum is close to being degenerate while exhibiting non-zero mass differences together with not too small mixing matrix elements.

We now turn to the contributions mediated by charged scalars, \textit{i.e.}\ the last two diagrams in Fig.~\ref{fig:LFV_muegAll}. In this case, the amplitudes can be written in a compact from once the light-heavy neutrino mixing matrix elements $\mathcal{V}^{\nu N}_{\alpha i}$ and $\mathcal{V}^{N \nu}_{\alpha i}$ are neglected,
\begin{equation}\begin{split}
    \mathcal{A}_L^{H_1^\pm} =&\ \tfrac{e m_\mu}{8 \pi^2} \tfrac{Y_L^{\mu \ast} Y_L^e}{M^2_{H_1^\pm}} \tfrac{k^2}{(k^2 + v_L^2)} \sum_i  \mathcal{V}^{N N}_{e i} \mathcal{V}^{N N \ast}_{\mu i} F_2 \left( \tfrac{M^2_{N_i}}{M^2_{H_1^\pm}} \right),\\
   \mathcal{A}_R^{H_1^\pm} =&\ \tfrac{e m_\mu}{8 \pi^2} \tfrac{Y^\mu Y^{e\ast}}{M^2_{H_1^\pm}}  \tfrac{v_L^2}{(k^2 + v_L^2)} \sum_i  \mathcal{V}^{\nu \nu}_{e i} \mathcal{V}^{\nu \nu \ast}_{\mu i} F_2 \left( \tfrac{m^2_{\nu_i}}{M^2_{H_1^\pm}}\right), \\
    \mathcal{A}_L^{H_2^\pm} =&\ \tfrac{e m_\mu}{8 \pi^2} \tfrac{Y^{\mu \ast} Y^e}{M^2_{H_2^\pm}} \tfrac{v_R^2}{(k^2 + v_R^2)} \sum_i  \mathcal{V}^{n n}_{e i} \mathcal{V}^{n n \ast}_{\mu i} F_2 \left( \tfrac{m^2_{n_i}}{M^2_{H_2^\pm}}\right),\\
    \mathcal{A}_R^{H_2^\pm} =&\ \tfrac{e m_\mu}{8 \pi^2} \tfrac{Y_R^{\mu \ast} Y_R^e}{M^2_{H_2^\pm}}  \tfrac{k^2}{(k^2 + v_R^2)} \sum_i \mathcal{V}^{n n}_{e i} \mathcal{V}^{n n \ast}_{\mu i} F_2 \left( \tfrac{m^2_{n_i}}{M^2_{H_2^\pm}}\right),
\end{split}\end{equation}
where the scalar function $F_2(x)$ defined in the limit $m_e \ll m_\mu \ll M_{H_{1,2}^\pm},~ m_{n_i},~ M_{N_i}$ is given by
\begin{equation}
    F_2 (x) = \tfrac{1-6x+3x^2 + 2x^3 - 6x^2 \text{log}(x)}{6 (1-x)^4}.
\end{equation}
This function takes a simpler form in well-defined limits, 
\begin{equation}
   F_2(x \rightarrow \infty ) = \tfrac{1}{3x} , ~~~  F_2(x \rightarrow 0 ) = \tfrac{1}{6} - x, ~~~ F_2(x \rightarrow 1 ) = \tfrac{7}{60} - \tfrac{x}{30} \,,
\end{equation}
so that the amplitudes driven by charged scalar exchanges only play a role when $M_{H_1^\pm}\approx M_{N}$ and $M_{H_2^\pm}\approx m_{n}$. In this case, the left-handed form factors approximately read
\begin{equation}\label{eq:scotino_amplitudes_bis}\begin{split}
  \mathcal{A}_L^{H_1^\pm} \approx&\ \tfrac{1}{30}~ \tfrac{e m_\mu}{8 \pi^2} \tfrac{Y_L^{\mu \ast} Y_L^e}{M^4_{H_1^\pm}}~ \left(\Delta M^2_{N_{12}}~ \mathcal{V}^{N N}_{e 2} \mathcal{V}^{N N \ast}_{\mu 2}+ \Delta M^2_{N_{13}}~ \mathcal{V}^{N N}_{e 3} \mathcal{V}^{N N \ast}_{\mu 3}\right),\\
  \mathcal{A}_L^{H_2^\pm} \approx&\ \tfrac{1}{30}~\tfrac{e m_\mu}{8 \pi^2} \tfrac{Y^{\mu \ast} Y^e}{M^4_{H_2^\pm}} ~ \left(\Delta m^2_{n_{12}}~ \mathcal{V}^{nn}_{e 2} \mathcal{V}^{nn \ast}_{\mu 2}+ \Delta m^2_{n_{13}}~ \mathcal{V}^{nn}_{e 3} \mathcal{V}^{nn \ast}_{\mu 3}\right),
\end{split}\end{equation}
and we do not provide expressions for the right-handed form factors as they turn out be be negligible for the parameter space considered in our analysis.

\begin{figure}
\includegraphics[width=0.7\textwidth]{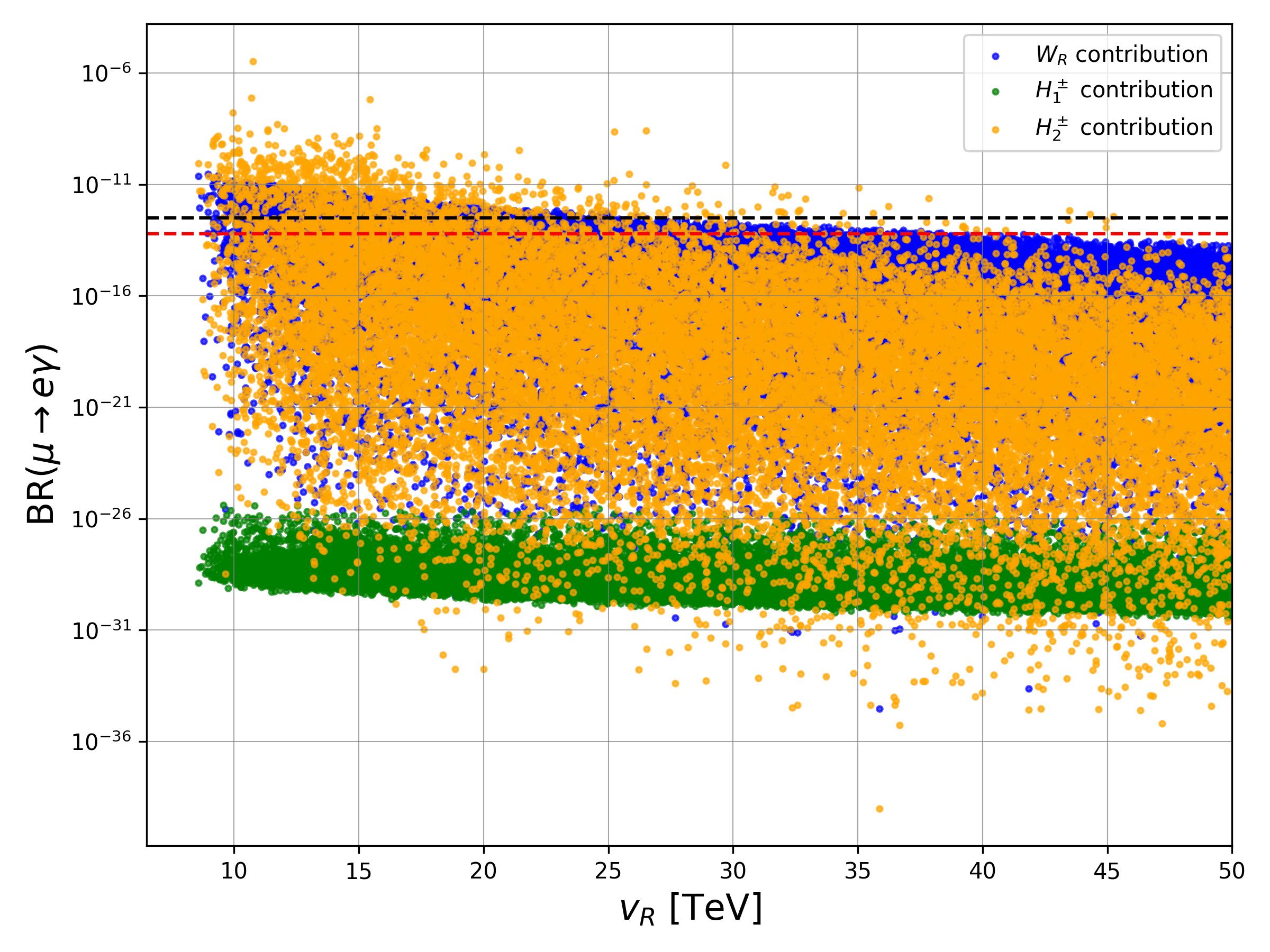}
\caption{Distributions of the contributions to BR$(\mu \to e \gamma)$ from $W_R/n_i$ (blue), $H_1^\pm/N_i$ (green) and $H_2^\pm/n_i$ (orange) exchanges, presented as a function of $v_R$. The horizontal dashed lines represent current bounds (black) and the future sensitivity (red) on this branching ratio, as expected from the MEG II experiment~\cite{Chiappini:2023luv, Baldini:2013ke}. Only predictions for scenarios in which $M_{Z^\prime} > $ 4.5~TeV are displayed. \label{fig:meg1}}
\end{figure}

In order to quantitatively assess the impact of these different contributions, we perform a numerical analysis exploring the different mass configurations sketched in Eqs.~\eqref{eq:scotino_amplitudes} and \eqref{eq:scotino_amplitudes_bis}. For this purpose, we implement a scan of the parameter space with the input parameters specified in Eqs.~\eqref{eq:input1} and \eqref{eq:inputNu}. With $\tan\beta$ varying in the given range, $v_L\ll k$ so that contributions from $A_ R^{H_1^\pm}$ are negligible, and $k\ll v_R$ so that $A_R^{H_2^\pm}$ is equally small, as already mentioned above. We can note that the impact of these contributions was already quite mild due to the smallness of $m_\nu$ and the corresponding approximate GIM suppression. Non-zero and significant contributions to the $\mu\to e \gamma$ LFV decay therefore arise predominantly from diagrams involving a left-handed muon. 

\begin{figure}
    \includegraphics[scale=0.46]{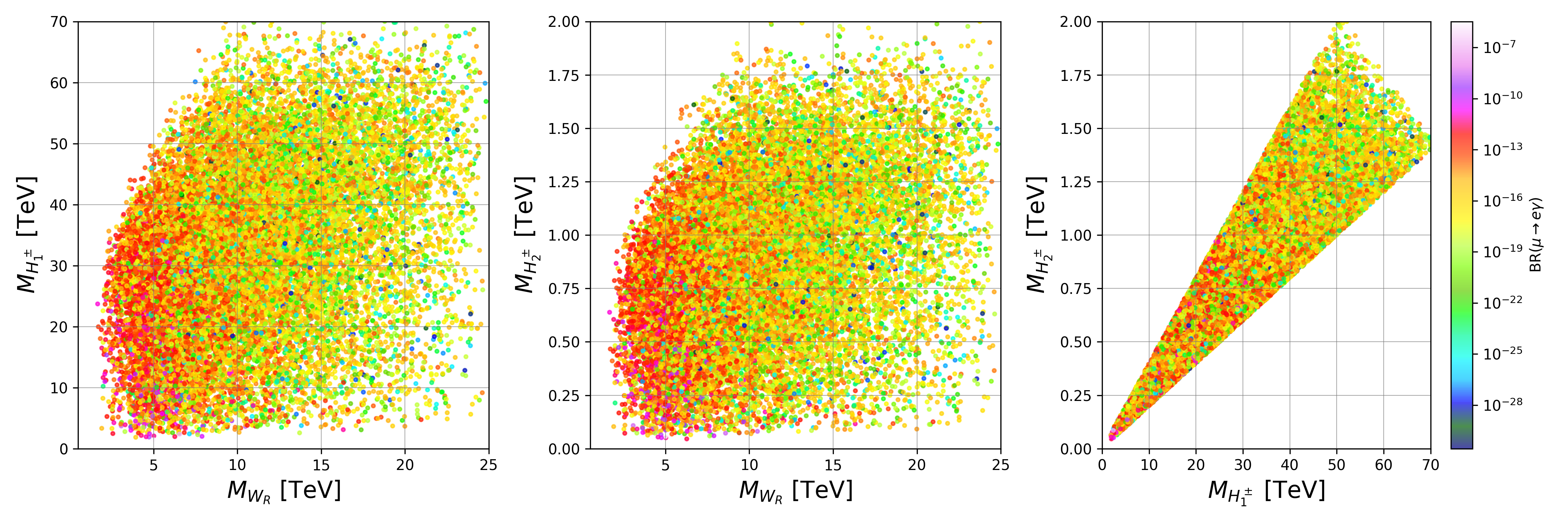}
    \caption{Predictions of the BR($\mu \rightarrow e \gamma$) branching ratio in the ALRM, shown within varied two-dimensional mass planes defined by the $W_R$, $H_1^\pm$ and $H_2^\pm$ masses.}
    \label{fig:enter-label}
\end{figure}

In Fig.~\ref{fig:meg1} we present predictions for the various contributions to BR$(\mu \to e \gamma)$, while separating the individual channels, and we depict their dependence on $v_R$. From all scanned points, we only retain those where the $Z^\prime$ boson is heavy enough to evade collider bounds, \textit{i.e.}\ with $M_{Z^\prime} > 4.5$~TeV. Contributions from $W_R$/scotino exchanges, $H_1^\pm$/heavy neutrino exchanges and $H_2^\pm$/scotino exchanges are shown in blue, green and orange, respectively. Given that $M^2_{H_1^\pm} \sim \tan\beta M^2_{H_2^\pm}$ for the scanned region of the parameter space, the first Higgs mass $M_{H_1^\pm}$ is always about an order or magnitude larger than the mass of the second Higgs state $M_{H_2^\pm}$. Therefore, contributions featuring $H_1^\pm$ mediation are expected to be much smaller than those mediated by $H_2^\pm$ exchanges. This is clearly illustrated in Fig.~\ref{fig:meg1}, which shows that $H_2^\pm$ contributions to  BR$(\mu \to e \gamma)$ can often be a few orders of magnitude larger than those involving $H_1^\pm$. Furthermore, despite the very large $ W_R$ boson mass, the associated contributions are always large by virtue of the couplings involved and often of similar size as those emerging from exchanges of the much lighter $H_2^\pm$ Higgs boson. We recall that throughout our parameter scan, the charged scalar $H_2^\pm$ is often found to be light, with a mass being one order of magnitude smaller than that of the $W_R$ boson. The charged Higgs contribution to the branching ratio BR$(\mu \to e \gamma)$ is consequently relatively enhanced. However, the $SU(2)_{R'}$ gauge coupling is always stronger than the small neutrino/scotino Yukawa couplings relevant for the charged Higgs contributions in the model considered. These $W_R$-boson exchange contributions are thus often by themselves sufficient to exclude a given scenario (as are those mediated by the second charged Higgs), or will soon be, as testified by the black and the red dashed lines representing the current and future bounds on the $\mu \to e \gamma$ branching ratio from the MEG~II experiment~\cite{Chiappini:2023luv, Baldini:2013ke}. 

In Fig.~\ref{fig:enter-label}, we display predictions (summed over all diagrams) for the (total) branching ratio BR$(\mu \to e \gamma)$ through a colour code, all scanned points being projected in three different two-dimensional mass planes. In the left and central panels of the figure, we consider the $(M_{W_R}, M_{H_1^\pm})$ and $(M_{W_R}, M_{H_2^\pm})$ mass planes respectively, while in the rightmost panel we present the results in the $(M_{H_1^\pm}, M_{H_2^\pm})$ plane. As just mentioned, for the whole range of values spanned by the $v_R$ parameter in our scan, diagrams featuring $W_R$ and $H_2^\pm$ boson exchanges yield the dominant contributions to the rare branching ratio, and those contributions are generally of the same order. The main difference arises from the masses of the two bosons, which is highlighted in the different panels of Fig.~\ref{fig:enter-label}. For scenarios with a light $W_R$ state with $M_{W_R} \lesssim 10$~TeV, BR$(\mu \to e \gamma)$ is generally the largest with branching ratio values often greater than $10^{-12}$, regardless of the mass of the second charge Higgs that could be equally below or above 1~TeV. For heavier $W_R$ boson with mass larger than 10~TeV, the branching ratio drops by several orders of magnitude. The connection between the mass of the second charged Higgs and that of the $W_R$ boson prevents the $H_2^\pm$ boson contributions to be large enough to compensate for this drop. Finally, we recall that the charged Higgs masses are strongly correlated, as mentioned above, and as shown in the right panel of the figure.

\begin{figure}
    \includegraphics[width=0.6\columnwidth]{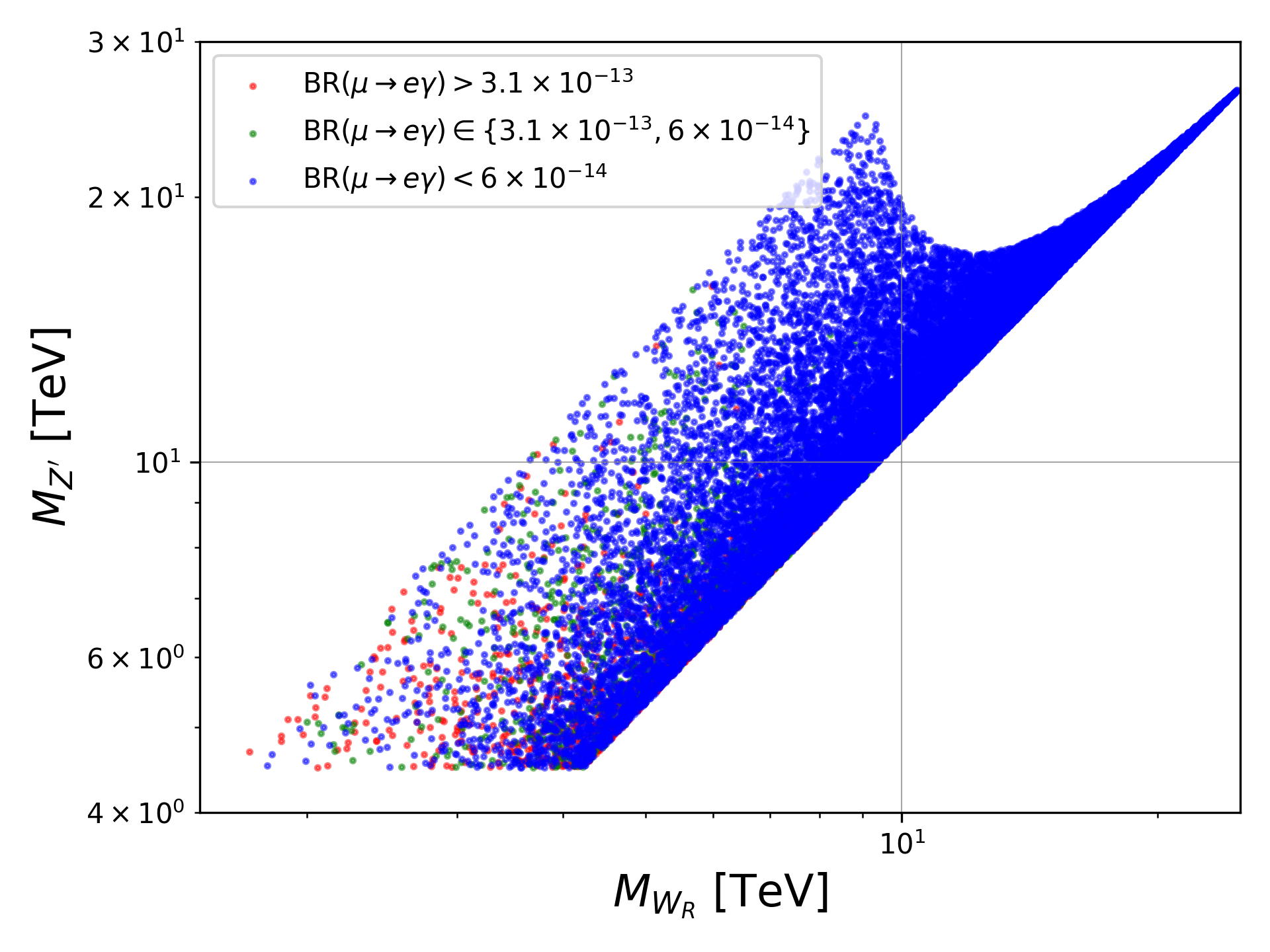}
    \caption{Allowed regions of the scanned parameter space projected in the $(M_{W_R}, M_{Z'})$ plane. We distinguish scenarios only allowed by current LHC bounds on the $Z'$ boson with $M_{Z^\prime} > 4.5$~TeV but excluded by current MEG~II measurement (red), scenarios that feature BR($\mu \rightarrow e \gamma$) $\in \lbrace 3.1\times 10^{-13}, 6 \times 10^{-14} \rbrace$ (green), and finally scenarios such that BR($\mu \rightarrow e \gamma$) $< 6\times 10^{-14}$ (blue).}
    \label{fig:MWR_MZR}
\end{figure}
We close this section by presenting in Fig.~\ref{fig:MWR_MZR}, the regions of the parameter space allowed by current LHC bounds on the ALRM $Z'$ state, and that could be further restricted by present bounds and future sensitivity on the rare branching ratio BR($\mu \rightarrow e \gamma$). All scanned scenarios satisfying LHC bounds on the $Z'$ boson (\textit{i.e.}, with $M_{Z^\prime} > 4.5$~TeV) are projected in the two-dimensional mass plane $(M_{W_R}, M_{Z^\prime})$. Such a representation allows us to determine, by means of the relationship between the extra gauge boson masses, how constraints on the $W_R$ boson from rare $\mu\to e\gamma$ decays could provide information on bounds on the neutral $Z'$ boson. All the points that are excluded by current MEG~II measurement on the branching ratio BR$(\mu \to e \gamma)$ are shown in red, whilst those which are sensitive to MEG~II present bound but will be disfavoured in the near future are shown in green. Finally, blue points refer to scenarios featuring a too small BR$(\mu \to e \gamma)$ value to be reachable by any planned experiment. We can infer from this figure that indirect LFV constraints have the potential to restrict the viable regions of the parameter space further, or equivalently that constraints could be extracted on the $W_R$ mass from improved measurements of the rare branching ratio BR$(\mu \rightarrow e \gamma)$. Subsequently, there is a potential to increase mass bounds on the ALRM $Z^\prime$ boson beyond those originating from the LHC, owing to the band-like structure which appears in the plane and that is due to varying the $g_R$ coupling in our scan and the sensitivity to BR$(\mu \to e \gamma)$. Finally, we emphasise that the specific shape of the upper-right boundary of the parameter space region populated by the points directly arises from the quadratic tree-level relationship between the $Z^\prime$ and $W_R$ masses, combined with the LHC mass limits imposed by $Z^\prime$ boson searches at the LHC.

\subsection{\texorpdfstring{$\mu-e$}{mu-e} conversion in nuclei}\label{sec:muconv}
In this section we analyse the impact of a typical ARLM spectrum on $\mu-e$ conversion in nuclei, which consists in one of the most promising experimental projects to test lepton flavour violation in the future. It has therefore attracted a lot of phenomenological interest for more than 50 years~\cite{Weinberg:1959zz, Marciano:1977cj, Czarnecki:1998iz, Kitano:2002mt, Cirigliano:2004mv, Cirigliano:2009bz, Borrel:2024ylg, Haxton:2024ecp, Haxton:2024lyc}. The summary of the current experimental status on the conversion rate for $\mu-e$ processes in aluminium, titanium and gold is presented in the second column of Table~\ref{tab:LFV}. Furthermore, these constraints are expected to be improved by several orders of magnitude in the near future, as shown in the third column of the table that collects expectations of experiments such as DeeMe at J-PARC~\cite{Natori:2018wcm} with its expected sensitivity of $\mathcal{O}(10^{-14})$, Mu2e at Fermilab \cite{DiFalco:2024bdw} with an expected sensitivity of $\mathcal{O}(10^{-17})$, and COMET at J-PARC \cite{Fujii:2023vgo} with a $\mathcal{O}(10^{-15})$ sensitivity in its first phase and $\mathcal{O}(10^{-17})$ in its second phase. These low bounds make $\mu-e$ conversions very attractive for constraining models featuring non-standard lepton-flavor-violating interactions like the ALRM.

In nuclei, $\mu - e$ conversions are traditionally studied within an effective field theory approach~\cite{Kuno:1996kv,Kitano:2002mt}. This effective description relies on a parton-level Lagrangian embedding higher-dimensional interactions between light quark ($q = u, d, s$), electron ($e$) and muon ($\mu$) fields,
\begin{align}
    - \mathcal{L}_{\text{eff}}^{(q)} & \subset - \sqrt{2} G_F \Biggl[ \sum_{X=L,R} \Biggl ( C_{DX} m_\mu (\bar{e} \sigma^{\alpha \beta} P_X \mu) F_{\alpha \beta} + C_{GX} \frac{\alpha_s}{4\pi}\frac{m_\mu}{v^2} (\bar{e} P_X \mu) G_{a \alpha \beta} G^{a \alpha \beta} \Biggr) \nonumber \\
    &\qquad + \sum_{q=u,d,s}\ \sum_{X, Y = L, R} \Biggl( C_{VXY}^{(q)} ( \bar{e} \gamma^\alpha P_X \mu ) (\bar{q} \gamma_\alpha P_Y q) + C_{SXY}^{(q)} ( \bar{e} P_X \mu ) (\bar{q} P_Y q) \Biggr ) + \mathrm{H.c.} \Biggr]\,.
    \label{eq:Lmue}
\end{align}
In this Lagrangian, the Wilson coefficients $C_{DX}$, $C_{GX}$, $C_{VXY}^{(q)}$ and $C_{SXY}^{(q)}$ encode the UV dynamics of the theory (with $X,Y = L,R$), and they are thus dependent on the model considered. They can be related in a second step to coefficients encoding the nucleon-level dynamics. To bridge the gap between the parton and nucleon regimes, we follow the methodology outlined in Refs.~\cite{Cirigliano:2009bz, Kitano:2002mt}. We extract proton-level and neutron-level effective coefficients,
\begin{equation}\begin{split}
    \widetilde{C}_{VX}^{(p)} =&\ \sum_{q_k = u, d, s} \sum_{Y = L,R} C_{VXY}^{(q)}  f_{Vp}^{(q_k)},\\
    \widetilde{C}_{VX}^{(n)} =&\ \sum_{q_k = u, d, s} \sum_{Y = L,R} C_{VXY}^{(q)}  f_{Vn}^{(q_k)}, \\
    \widetilde{C}_{SX}^{(p)} =&\ \sum_{q_K = u,d,s} \sum_{Y = L,R} \dfrac{m_{p}}{m_{q}} C^{(q)}_{SXY} f^{(q_k)}_{Sp} + \dfrac{m_\mu m_{p}}{4\pi v^2} C_{G_X} f_{Gp},\\
   \widetilde{C}_{SX}^{(n)} = &\ \sum_{q_K = u,d,s} \sum_{Y = L,R} \dfrac{m_{p}}{m_{q}} C^{(q)}_{SXY} f^{(q_k)}_{Sn} + \dfrac{m_\mu m_{p}}{4\pi v^2} C_{G_X} f_{Gn},
\end{split}\end{equation}
where the various $f$ coefficients are the conventional nucleon, scalar and gluonic form factors~\cite{Crivellin:2017rmk, Crivellin:2014cta},
\begin{equation}\begin{split}
    & f_{Vp}^{(u)} = 2,\qquad f_{Vp}^{(d)} = 1,\qquad f_{Vp}^{(s)} = 0, \qquad 
       f_{Vn}^{(u)} = 1,\qquad f_{Vn}^{(d)} = 2,\qquad f_{Vp}^{(s)} = 0, \\
    & f_{Sp}^{(u)} = (20.8 \pm 1.5) \times 10^{-3}, \qquad f_{Sp}^{(d)} = (41.1 \pm 2.8) \times 10^{-3}, \qquad f_{Sp}^{(s)} = (53.0 \pm 27) \times 10^{-3}, \\
    & f_{Sn}^{(u)} = (18.9 \pm 1.4) \times 10^{-3}, \qquad f_{Sn}^{(u)} = (45.1 \pm 2.7) \times 10^{-3}, \qquad f_{Sn}^{(s)} = (53.0 \pm 27) \times 10^{-3}, \\
    & f_{G p(n)} = - \dfrac{8\pi}{9} \Big[ 1- \sum_{q = u,d,s} f_{S p(n)}^{(q)} \Big].
\end{split}\label{eq:formfactors}\end{equation}
The $\mu \rightarrow e$ conversion rate normalised to the muon capture rate ${\Gamma_{\text{capt}}}$ in a specific nucleus $N$ is then given by \cite{Kitano:2002mt}
\begin{align}
    \text{CR} (\mu - e) (N) = \tfrac{m_\mu^5}{4 \Gamma_{\text{capt}} (N)} \left| C_{DL} D_N + 2 \Big[ \widetilde{C}^{VR}_{(p)} V_N^{(p)} + \widetilde{C}^{SL}_{(p)}  S^{(p)}_N + (p \to n) \Big] \right|^2  + (L \leftrightarrow R) , 
\label{eq:gammaconv}\end{align}
where the dimensionless integrals $D_N$ and $V_N^{(q)}$ represent the overlap between the electron and muon wave functions. Explicit expressions for these integrals can be found in Refs.~\cite{Czarnecki:1998iz, Kitano:2002mt}, and the muon capture rate ${\Gamma_{\text{capt}} (N)} $ by a nucleus $N$ with atomic and mass numbers $Z$ and $A$ is related to the cross section associated with the process
\begin{equation}
\mu^-+(Z,A) \to \nu_\mu +(Z-1,A),~~~~~~{\rm or~equivalently}~~~~~\mu^- + p \to \nu_\mu + n.
\end{equation}
Numerical values for all relevant nuclear physics inputs appearing in Eq.~\eqref{eq:gammaconv} are available from~\cite{Kitano:2002mt, Heeck:2022wer, Suzuki:1987jf}, and we collect them in Table~\ref{tab:muevalues}.

\begin{table} \centering\setlength\tabcolsep{8pt} \renewcommand{\arraystretch}{1.3}
   \begin{tabular}{c|ccccc|c}
     Isotope & $D_N$ \cite{Kitano:2002mt} &  $V^{(p)}$ \cite{Kitano:2002mt} &  $V^{(n)}$ \cite{Kitano:2002mt} & $S^{(p)}$ \cite{Heeck:2022wer} & $S^{(n)}$ \cite{Heeck:2022wer} & $\Gamma_{\text{capt}} (10^6$ s$^{-1})$ \cite{Suzuki:1987jf} \\
     \hline
     $^{27}_{13}$Al & 0.0359 & 0.0165  & 0.0178 & 0.0159 & 0.0172 & 0.69 \\
     $^{48}_{22}$Ti & 0.0859 & 0.0407 & 0.0481 & 0.0379 & 0.0448 & 2.59\\
     $^{127}_{79}$Au & 0.1660 & 0.0866 & 0.1290 & 0.0523 & 0.0781 & 13.07 \\
   \end{tabular}
\caption{Numerical values for the nuclear physics inputs and muon capture rates in different isotopes, that are relevant for the calculation of $\mu-e$ conversion rates performed in this work. }
\label{tab:muevalues}
\end{table}

\begin{figure}
    \centering
    \includegraphics[scale=0.35]{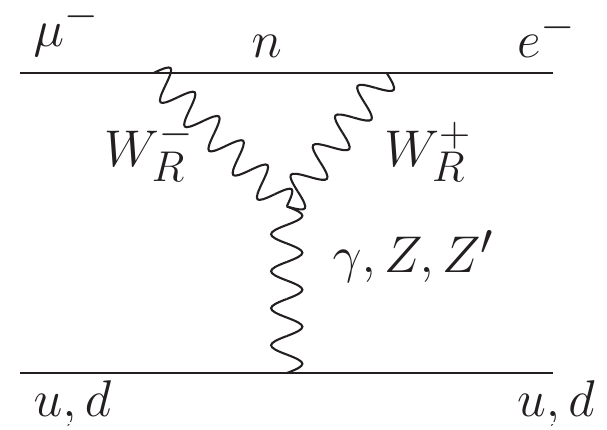}\qquad
    \includegraphics[scale=0.35]{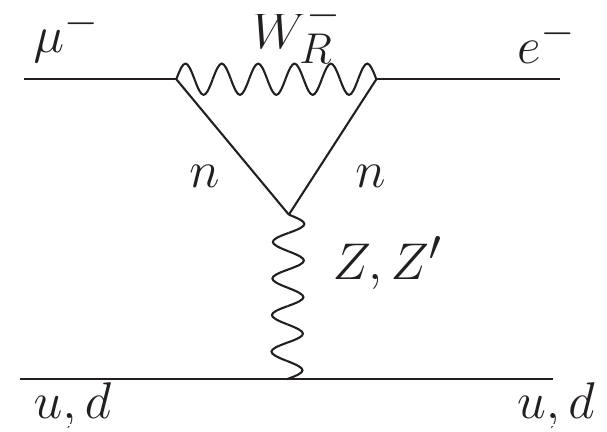}\qquad
    \includegraphics[scale=0.35]{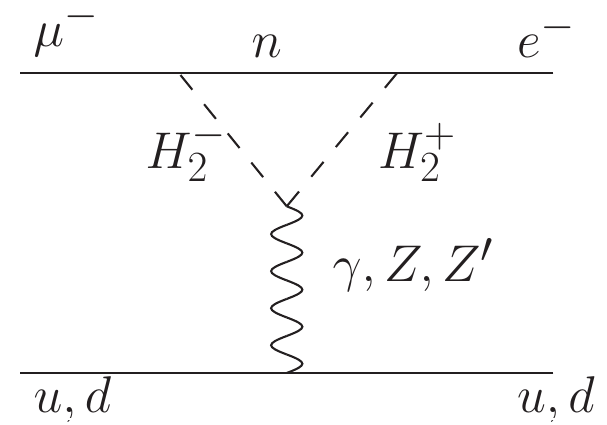}\qquad
    \includegraphics[scale=0.35]{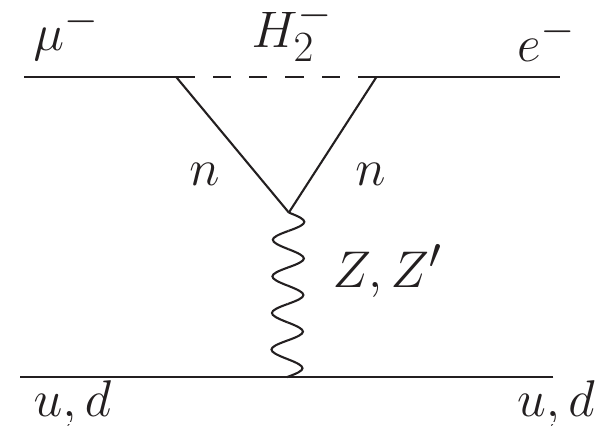}\qquad
    \includegraphics[scale=0.35]{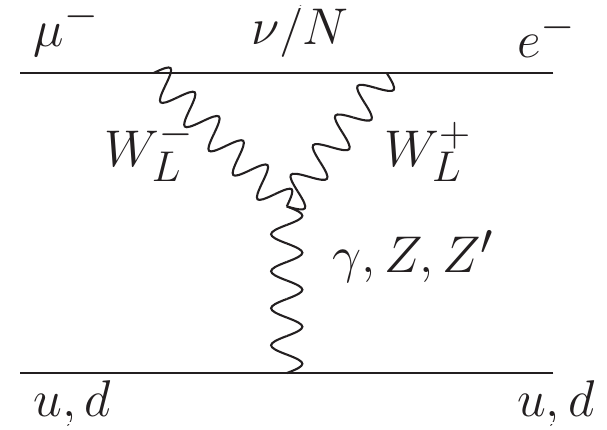}\qquad
    \includegraphics[scale=0.35]{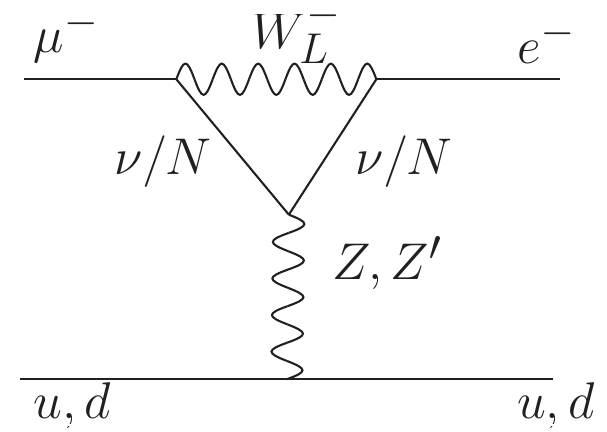}\qquad
    \includegraphics[scale=0.35]{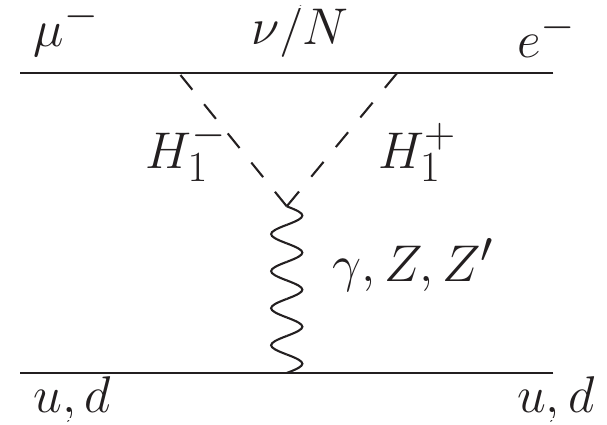}\qquad
    \includegraphics[scale=0.35]{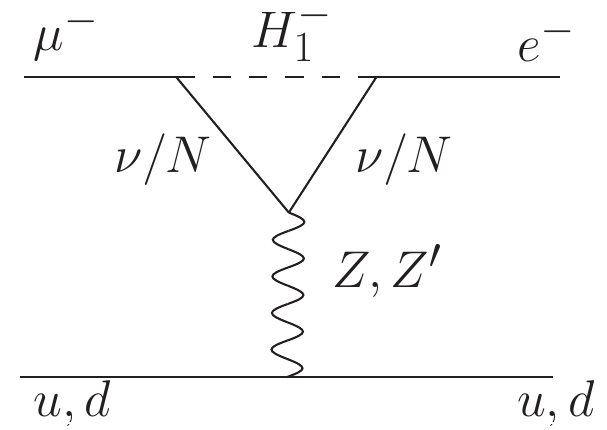}
    \caption{Feynman diagrams driving $\mu-e$ conversion processes in nuclei in ALRM scenarios. The two top (bottom) leftmost diagrams represent contributions from $W_R^\pm$ ($W_L^\pm$) -boson exchanges, while the two top (bottom) rightmost diagrams are related to contributions from $H_2^\pm$ ($H_1^\pm$) exchanges.
    \label{fig:crmue}}
\end{figure}

In the ALRM and at one loop, $\mu-e$ conversions solely arise from triangle contributions, since any potential box diagram involves joint $W_L$-boson and $H_1^\pm$-boson exchanges and is therefore vanishing. The relevant Feynman diagrams are shown in Fig.~\ref{fig:crmue}. In the following, we mainly focus on $t$-channel photonic exchanges, because diagrams featuring mediation via a heavy neutral gauge boson are suppressed by a massive vector propagator, at least when compared with diagrams including a massless photon propagator. We are thus left to estimate the impact of few remaining diagrams on the Wilson coefficients appearing in the Lagrangian of Eq.~\eqref{eq:Lmue}.

We perform our computations in the elastic limit, as typical momentum transfers given by $t=(p_{(\mu)}-p_{(e)})^2$ are of the order of $m_\mu^2$. In contrast to the computations relevant for $\mu \to e \gamma$ decays that we have addressed in the previous section, the photon appearing in the muon conversion diagrams is off-shell. The limit $t\to 0$ can thus be safely taken, especially as $m_\mu^2 \sim 0$ relatively to the large masses of the charged bosons running into the loops. Considering this limit as well as negligible light-heavy neutrino mixing ($V^{\nu N}, V^{N\nu}$); several Wilson coefficients of Eq.~\eqref{eq:Lmue} are modified by ALRM effects,
\begin{align}
     & C_{DL} (H_1^\pm) = \tfrac{e}{16 \pi^2} \tfrac{1}{M_{H_1^\pm}^2} \left[Y^{\mu \ast}_L Y^e_L \tfrac{k^2}{k^2 + v_L^2}\right]\sum_i V_{\mu i}^{NN \ast} V_{\mu i}^{NN} F_2 \left(\tfrac{M_{N_i}^2}{M_{H_1^\pm}^2}\right), \nonumber \\
    & C_{DR} (H_1^\pm) = \tfrac{e}{16 \pi^2} \tfrac{1}{M_{H_1^\pm}^2} \left[Y^{\mu \ast} Y^e \tfrac{v_L^2}{k^2 + v_L^2}\right]\sum_i V_{\mu i}^{\nu\nu \ast} V_{\mu i}^{\nu\nu} F_2 \left(\tfrac{m_{\nu_i}^2}{M_{H_1^\pm}^2}\right), \nonumber \\
    & C_{DL}(H_2^\pm) = \tfrac{e}{16 \pi^2} \tfrac{1}{M_{H_2^\pm}^2} \left[ Y^{\mu \ast} Y^e \tfrac{v_R^2}{(k^2 + v_R^2)} \right] \sum_i \mathcal{V}_{\mu i}^{nn \ast} \mathcal{V}_{ei}^{nn} F_2 \left( \tfrac{M_{n_i}^2}{M_{H_2^\pm}^2} \right),\nonumber \\
    & C_{DR}(H_2^\pm) = \tfrac{e}{16 \pi^2} \tfrac{1}{M_{H_2^\pm}^2} \left[ Y^{\mu \ast}_R Y^e_R \tfrac{k^2}{(k^2 + v_R^2)} \right] \sum_i \mathcal{V}_{\mu i}^{nn \ast} \mathcal{V}_{ei}^{nn} F_2 \left( \tfrac{M_{n_i}^2}{M_{H_2^\pm}^2} \right),\nonumber \\
    & C_{DL}(W_L^\pm) = \tfrac{e}{64\pi^2} \tfrac{1}{M_{W_L^\pm}^2}\tfrac{g_L^2}{4} \sum_i V^{\nu\nu \ast}_{\mu i} V^{\nu\nu}_{e i} F_1 \left(\tfrac{m_{\nu_i}^2}{M_{W_L^\pm}^2}\right), \nonumber \\
    & C_{DR}(W_R^\pm) = \tfrac{e}{64 \pi^2} \tfrac{1}{M_{W_R^\pm}^2} \tfrac{g_R^2}{4} \sum_i \mathcal{V}_{\mu i}^{nn \ast} \mathcal{V}_{ei}^{nn} F_1 \left( \tfrac{M_{n_i}^2}{M_{W_R^\pm}^2} \right),
\end{align}
the contributions to the other coefficients being either exactly zero (like for the vector coefficient $C_{VRR}^{(q)}$) or suppressed by the small size of the neutrino masses and related couplings (like for $C_{VLL}^{(q)}$, $C_{VLR}^{(q)}$ and $C_{VRL}^{(q)}$), heavy neutral scalar propagators (like for scalar coefficients) or the smallness of the scalar form factors $f_Si$ in Eq.~\eqref{eq:formfactors}, compared to the order of magnitude of the other form factors. We refer to the discussion in Ref.~\cite{FileviezPerez:2017zwm} for comprehensive details on the different contributions to muon conversions in left-right models. Among all the contributions, the triangle diagrams exhibiting $W_L$ and $H_1^\pm$ boson exchanges are negligible due to GIM suppression and the smallness of the light-neutrino masses ($m_{\nu}\ll M_{W_L}, M_{H_1^\pm}$). Moreover, contributions arising from the exchange of $H_1^\pm$ charged Higgs bosons are significantly further suppressed due to its mass. The large mass of the $H_1^\pm$ boson indeed leads to suppression in the corresponding propagators, thus reducing the overall impact of these contributions on the relevant processes. Therefore, the only relevant contributions come from the Feynman diagrams mediated by $W_R$-boson and $H_2^\pm$-boson exchanges.

\begin{figure}
    \centering
    \includegraphics[scale=0.46]{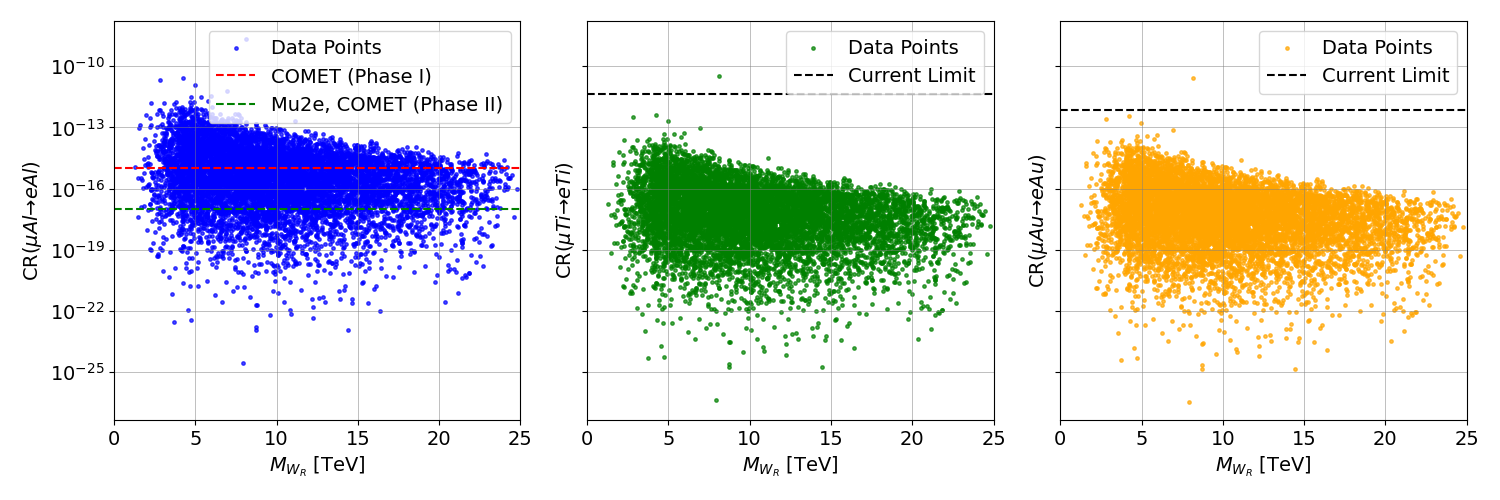}
    \caption{Muon capture rates in Al (left), Ti (centre) and Au (right) as returned by our scan of the ALRM parameter space, and displayed as a function of the $W_R$ boson mass. The sensitivities of several current and future experiments are indicated through dashed horizontal lines. Here we have imposed the experimental constraint of $M_{Z^{\prime}} > 4.5$~TeV while scanning the parameter space. \label{fig:LFV_mueg}}
\end{figure}

In Fig.~\ref{fig:LFV_mueg} we compare our predictions to bounds on $\mu - e$ conversion stemming from current measurements, and to expectations for the sensitivity of several future experiments. We scan over the parameter space following the inputs specified in Eqs.~\eqref{eq:input1} and \eqref{eq:inputNu}, and for each point we evaluate the muon capture rate in aluminium (left), titanium (centre) and gold (right) as a function of $W_R$ mass. We observe that mild bounds on the ALRM parameter space already exist, from achieved measurements, but that future experiments have the potential to strongly restrict the phenomenologically viable part of the parameter space. This is particularly pronounced for muon capture in aluminium, where future measurements have the potential to yield stringent constraints on ALRM scenarios with a charged vector boson mass $M_{W_R} \ge 1300$ GeV and drastically reduce the range of acceptable possibilities. 

\begin{figure}
\centering
\includegraphics[width=0.7\linewidth]{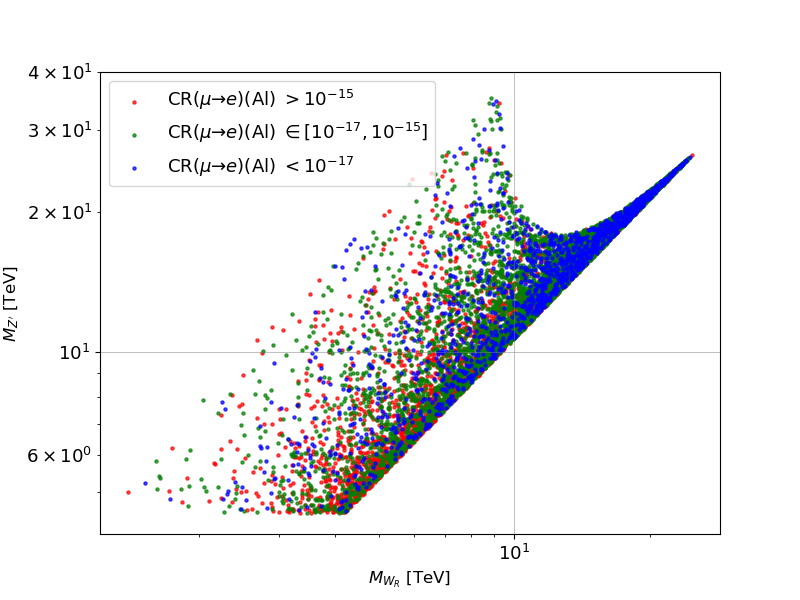}
\caption{Muon capture rates in Al, projected in the two-dimensional mass plane $(M_{W_R}, M_{Z'})$. We display predictions for all scenarios explored in our scan, the colour code indicating whether a given scenario has the potential to be probed in the future at COMET-I (red), and at Mu2e and COMET-II (green), or the predictions are such that the scenario is not reachable at any future planned experiment (blue). Here we have imposed the experimental constraint of $M_{Z^{\prime}} > 4.5$~TeV while scanning the parameter space.}
\label{fig:test}
\end{figure}

Finally, we explore the potential of using the existing correlation between the $W_R$-boson and $Z'$-boson masses to indirectly constrain the properties of ALRM $Z'$ boson via $W_R$ contributions to $\mu-e$ transitions in nuclei. To this aim, we project in  Fig.~\ref{fig:test} all points of our scan in the $(M_{W_R}, M_{Z'})$ mass plane, focusing on conversions in aluminum, as only this nucleus has the potential to constrain the ALRM (as shown in Fig.~\ref{fig:LFV_mueg}). Our findings show that a large class of scenarios featuring a $Z'$ boson with mass lying between 5 and 35~TeV have the potential to be excluded for $M_{W_R}\lesssim 12$~TeV (and even in the heaviest $Z'$ setups). However, many scenarios will still remain beyond the reach of present and future experiments for all values of the $W_R$ boson mass (and all corresponding values of the $Z'$-boson mass), the associated contributions being not large enough to provide any substantial enhancement. Consequently, while indirect $\mu-e$ conversion constraints will strengthen our knowledge of which ALRM configuration will remain phenomenologically acceptable in a close future, existing conspiracy between the numerous free parameters of the model (in particular $v_R$ and $g_R$) will hinder our ability to draw a generic conclusion on the possible masses for the extra charged and neutral gauge bosons.

\section{Conclusion}\label{sec:conclusion}
We explored an alternative version of the left-right symmetric model emerging from a grand unified theory that embeds a gauge symmetry based on the exceptional $E_6$ group. The breaking of this symmetry group to the electroweak symmetry group yields an intermediate left-right symmetry pairing the SM fermions with exotic down-type quark and neutrino states, the latter being dubbed scotinos. Moreover, the symmetry breaking pattern originates from the presence of an enriched Higgs sector featuring several neutral and charged physical Higgs bosons. In this work, we assess how this spectrum of extra fields has the potential to impact several leptonic observables, and how existing and future associated measurements have the potential to constrain the model. Very interestingly, in the ALRM the $W_R$ boson is only constrained at colliders indirectly, through the relationship between its mass and the extra neutral boson $Z^\prime$. Conversely, we investigate how indirect probes from leptonic observables on the $W_R$ boson could bring complementary information. 

We chose the focus first on the anomalous magnetic moment of the muon, and we demonstrated by means of a vast scan of the model parameter space that ALRM scenarios have little chance to contribute largely enough to explain the longstanding deviations between the SM predictions and measurements. We calculated one-loop ALRM contributions, that are suppressed due to the relatively high masses of new bosons as compared to the muon mass, as well as the two-loop ones, that suffer from a conspiracy involving the smallness of the relevant Yukawa couplings and the masses of the new fermions running into the loops to keep  the full contribution small across the entire parameter space. The negligible ALRM impact on the anomalous magnetic moment of the muon, however, gels well with the most recent lattice results, that advocate agreement between SM and data.

On the other hand, we studied two classes of lepton-flavour-violating processes, namely muon-to-electron conversion in nuclei and muon decays to an electron-photon system. We determined how predictions are impacted in the ALRM at one loop. We perform a general scan of the parameter space, and demonstrated the potential of future related measurements to constrain the masses of the charged Higgs and gauge bosons of the model. Both present and future experimental data have hence the potential to restrict ALRM scenarios featuring a $W_R$-boson mass larger than 1.3 TeV (from $\mu - e$ conversion in Al), and a charged Higgs masses for scenarios featuring both light and heavy states.

These results are complementary to the mass bounds expected to be obtained at colliders, where the $W_R$ and charged Higgs bosons cannot be directly produced. Present bounds and future sensitivity from MEG II can constrain the $Z'$ mass parameter space more tightly as compared to LHC bounds, but a quantitative limit is difficult to state since the $W_R$ and $Z'$ masses depend on both $v_R$ and $g_R$ parameters which can be varied independently. Hence here we have performed calculations offering the possibility to alternatively constrain the model, even in case where the masses are much larger than existing collider energies.

\section*{Acknowledgements}  
The work of MF is funded in part by NSERC  under grant number SAP105354, and the one of BF has received partial support from the French \emph{Agence Nationale de la Recherche} (ANR) through Grant ANR-21-CE31-0013 (DMwithLLPatLHC).  The work of SKG and PP has been supported by SERB, DST, India through grants TAR/2023/000116 and CRG/2022/002670 respectively. SKG acknowledges Manipal Centre for Natural Sciences, Centre of Excellence, Manipal Academy of Higher Education (MAHE) for facilities and support. CM acknowledges the support of the Royal Society, UK, through the Newton International Fellowship (grant number NIF$\backslash$R1$\backslash$221737). The work of SS is supported in part under the US Department of Energy contract DE-SC0011095.

\appendix 
\section{ALRM Couplings to Scalars and Vectors} 
\label{sec:appendix}

\begin{table}
  \centering \setlength\tabcolsep{8pt} \renewcommand{\arraystretch}{1.05}
  \resizebox{.98\columnwidth}{!}{\begin{tabular}{ccc}
    Interaction & Vector $(g_V)$ & Axial-vector $(g_A)$ \\ \hline
    $\overline{\ell}_i \ell_i Z $ & $i [-\tfrac{e g_L}{2 g_Y} + \tfrac{3e g_Y}{2 g_L}]$ & $i[\tfrac{e g_L}{2 g_Y} + \tfrac{e g_Y}{2 g_L}]$ \\
    $\overline{\ell}_i \ell_i Z^\prime $ & $i \tfrac{- g_R^2 + 3 g_Y^2}{2 \sqrt{g_R^2-g_Y^2}} $ & $ i \tfrac{- g_R^2 + g_Y^2}{2 \sqrt{g_R^2-g_Y^2}} $ \\
    $ \overline{\ell}_j \nu_i W_L^-$   & $i \tfrac{g_L}{2 \sqrt{2}} \mathcal{V}^{\nu \nu }_{j i}$  & $ -i \tfrac{g_L}{2 \sqrt{2}} \mathcal{V}^{\nu \nu }_{j i}$ \\ 
    $ \overline{\ell}_j N_i W_L^-$   & $i \tfrac{g_L}{2 \sqrt{2}} \mathcal{V}^{\nu N }_{j i}$  & $ -i \tfrac{g_L}{2 \sqrt{2}} \mathcal{V}^{\nu N }_{j i}$ \\ 
    $\overline{\ell}_j n_{i} W_R^-$   & $i \tfrac{g_R}{2 \sqrt{2}} \mathcal{V}^{ n n \ast}_{j i}$  & $ i\tfrac{g_R}{2\sqrt{2}} \mathcal{V}^{ n n \ast}_{j i}$ \\ 
  \end{tabular} \hspace{.3cm}
  \begin{tabular}{ccc}
    Interaction & Scalar $(g_S)$ & Pseudo-scalar $(g_P)$ \\ \hline
    $\overline{\ell}_i \ell_i h$ & $-i Y^\ell_{ii} U_{11}^H$ & 0 \\
    $\overline{\ell}_i \ell_i H_2^0$ & $-i Y^\ell_{ii} U_{12}^H$ & 0 \\
    $\overline{\ell}_i \ell_i H_3^0$ & $-i Y^\ell_{ii} U_{13}^H$ & 0 \\
    $\overline{\ell}_i \ell_i A_2^0$ & 0 & $ Y^\ell_{ii} U_{13}^A$ \\
    $\overline{\ell}_j \nu_i H_1^-$ & $ i \tfrac{Y^\ell_{kj} \mathcal{V}^{\nu \nu}_{ki} v_L + Y^{\ell}_{L, jk}  \mathcal{V}^{ \nu N}_{ki} k}{ \sqrt{k^2 + v_L^2}}$ & $i \tfrac{- Y^\ell_{kj} \mathcal{V}^{\nu \nu}_{ki} v_L + Y^{\ell}_{L,jk}  \mathcal{V}^{ \nu N}_{ki} k}{\sqrt{k^2 + v_L^2}}$ \\ 
    $\overline{\ell}_j N_i H_1^-$ & $i\tfrac{Y^\ell_{kj} \mathcal{V}^{N \nu}_{ki} v_L + Y^{\ell}_{L,jk}  \mathcal{V}^{N N}_{ki} k}{ \sqrt{k^2 + v_L^2}}$ & $i\tfrac{-Y^\ell_{kj} \mathcal{V}^{N \nu}_{ki} v_L + Y^{\ell}_{L,jk}  \mathcal{V}^{ N N}_{ki} k}{\sqrt{k^2 + v_L^2}}$ \\ 
    $\overline{\ell}_j n_i H_2^-$ & $i\tfrac{ \mathcal{V}^{nn}_{ki} (Y^\ell_{jk} v_R + Y^{\ell}_{R,jk} k)}{\sqrt{k^2 + v_R^2}}$ & $i\tfrac{ \mathcal{V}^{nn}_{ki} (Y^\ell_{jk} v_R - Y^{\ell}_{R,jk} k)}{\sqrt{k^2 + v_R^2}}$ \\ 
  \end{tabular}}
  \caption{FFV and FFS interactions of massive gauge and Higgs bosons with leptons, the coupling strengths being split into their vector/scalar ($g_V$/$g_A$, second column) and axial-vector/pseudu-scalar ($g_A$/$g_P$, third column) components. The indices $i$ and $j$ refer to generation indices.\label{tab:FFV_FFS}}
\end{table}

In this appendix, we list the relevant couplings used in the calculations presented in this paper, as extracted from the Lagrangian introduced in Section~\ref{sec:mod}. Table~\ref{tab:FFV_FFS} collects the fermion-fermion-vector (FFV) and fermion-fermion-scalar (FFS) couplings involving the lepton sector and entering the one-loop contributions to the observables considered. These couplings depend on the electromagnetic, $SU(2)_L$, $SU(2)_{R'}$ and hypercharge gauge couplings $e$, $g_L$, $g_R$ and $g_Y$, as well as on the Yukawa couplings, VEVs and mixing matrices defined in Section~\ref{sec:mod}.

\begin{table}
  \centering \setlength\tabcolsep{7pt} \renewcommand{\arraystretch}{0.95}
  \begin{tabular}{ccc}
    Interaction & Scalar ($g_S$) & Pseudo-scalar ($g_P$) \\ \hline
    $\overline{t} t h $ & $ -i Y^q_{33} U_{11}^H $ &  0\\ 
    $\overline{t} t H_2^0 $ & $ -i Y^q_{33} U_{12}^H $ &  0\\ 
    $\overline{t} t H_3^0 $ & $ -i Y^q_{33} U_{13}^H $ &  0\\ 
    $\overline{t} t A_2^0 $ & 0 &  $ - Y^q_{33} U_{13}^A $\\
  \end{tabular}\hspace{.5cm}
  \begin{tabular}{ccc}
    Interaction & Scalar ($g_S$) & Pseudo-scalar ($g_P$) \\ \hline
    $\overline{b} b h $ & $ -i Y^q_{L,33} U_{21}^H $ &  0 \\ 
    $\overline{b} b H_2^0 $ & $ -i Y^q_{L,33} U_{22}^H $ &  0 \\ 
    $\overline{b} b H_3^0 $ & $ - i Y^q_{L,33} U_{23}^H $ &  0 \\ 
    $\overline{b} b A_2^0 $ & 0 &  $ - Y^q_{L,33} U_{23}^A $ \\
  \end{tabular}
  \begin{tabular}{ccc}
    Interaction & Scalar ($g_S$) & Pseudo-scalar ($g_P$) \\ \hline
    $\overline{q}_{d,i}^\prime q_{d,i}^\prime h $ & $ -i Y^q_{R,ii} U_{31}^H $  & 0\\ 
    $\overline{q}_{d,i}^\prime q_{d,i}^\prime H_2^0 $ & $ -i Y^q_{R,ii} U_{32}^H $  & 0\\ 
    $\overline{q}_{d,i}^\prime q_{d,i}^\prime H_3^0 $ & $ -i Y^q_{R,ii} U_{33}^H $  & 0\\ 
    $\overline{q}_{d,i}^\prime q_{d,i}^\prime A_2^0 $ & 0 & $ - Y^q_{R,ii} U_{33}^A$\\
    \end{tabular}
    \caption{As in Table~\ref{tab:FFV_FFS}, but for the FFS interactions involving SM and exotic quarks and contributing at two loops to $\Delta a_\mu$. The index $i$ is a generation index.}
    \label{tab:2loop1}
\end{table}

In the rest of this section, we focus on the interactions relevant for the two-loop ALRM contributions to the anomalous magnetic moment of the muon, $\Delta a_\mu$. The couplings relevant for diagram~(1) in Fig.~\ref{fig:2loop1} include the FFS couplings of Table~\ref{tab:FFV_FFS} involving a neutral Higgs boson, as well as the FFS couplings to SM top and bottom quarks and to exotic $q_d' = d', s', b'$ quarks given in Table~\ref{tab:2loop1}. These quark couplings are again split in their scalar and pseudo-scalar components, and written in terms of the Yukawa couplings and mixing matrices introduced in Section~\ref{sec:mod}. Diagram~(2) involves additional trilinear scalar interactions (SSS) that we provide in Table~\ref{tab:2loop2}, and diagram~(3) involves the charged-current gauge interactions of the model's Higgs bosons (VVS) that we display in Table~\ref{tab:2loop3}, along with the lepton-scalar couplings already introduced in Table~\ref{tab:FFV_FFS}. Diagrams~(4a) and (4b) depend on the FFV and FFS couplings of~\ref{tab:FFV_FFS}, the SSS couplings of Table~\ref{tab:2loop2}, and the VSS couplings of Table~\ref{tab:2loop3}, while diagrams (5a) and (5b) involve the FFS and FFV couplings of Table~\ref{tab:FFV_FFS}, and the VVS and SSV couplings of Table~\ref{tab:2loop3}. Finally, diagrams (6a) and (6b) depends on the FFS and FFV couplings of Table~\ref{tab:FFV_FFS}, as well as on the charged-current FFS and FFV interactions with SM and exotic quarks of Table~\ref{tab:2loop6}. The latter are proportional to the SM and exotic CKM matrices $V_{\rm CKM}$ and $V'_{\rm CKM}$ driving the mixing of the SM down-type quarks and exotic quarks, respectively.

\begin{table}
  \centering\setlength\tabcolsep{10pt} \renewcommand{\arraystretch}{1.2}
  \begin{tabular}{cc}
    Interaction & Coupling\\ \hline 
    $H_1^+ H_1^- h $ & $ \tfrac{i[-4 k^3 U_{11}^H \alpha -4 v_L^3 U_{21}^H  \alpha - 4 v_L^2 v_R U_{31}^H \alpha -2 k v_L^2 U_{11}^H  \lambda_1 - 2 k^2 v_L U_{21}^H  \lambda_3 -2 k^2 v_R U_{31}^H  \lambda_4 +\sqrt{2} k v_L U_{31}^H \kappa]}{(k^2 + v_L^2)}$ \\ 
    $H_1^+ H_1^- H_2^0 $ & $ \tfrac{i[-4 k^3 U_{12}^H \alpha -4 v_L^3 U_{22}^H  \alpha - 4 v_L^2 v_R U_{32}^H \alpha -2 k v_L^2 U_{12}^H  \lambda_1 - 2 k^2 v_L U_{22}^H  \lambda_3 -2 k^2 v_R U_{32}^H  \lambda_4 + \sqrt{2} k v_L U_{32}^H \kappa]}{(k^2 + v_L^2)}$ \\ 
    $H_1^+ H_1^- H_3^0 $ & $ \tfrac{i[-4 k^3 U_{13}^H \alpha -4 v_L^3 U_{23}^H  \alpha - 4 v_L^2 v_R U_{33}^H \alpha -2 k v_L^2 U_{13}^H  \lambda_1 - 2 k^2 v_L U_{23}^H  \lambda_3 -2 k^2 v_R U_{33}^H  \lambda_4 +\sqrt{2} k v_L U_{33}^H \kappa]}{(k^2 + v_L^2)}$ \\ 
    $H_2^+ H_2^- h $ & $ \tfrac{i[- 4 k^3 U_{11}^H \alpha - 4 v_L v_R^2 U_{21}^H \alpha - 4 v_R^3 U_{31}^H \alpha - 2 k v_R^2 U_{11}^H  \lambda_1 - 2 k^2 v_R U_{31}^H \lambda_3 - 2 k^2 v_L U_{21}^H \lambda_4 + \sqrt{2} k v_R U_{21}^H \kappa]}{(k^2 + v_R^2)}$ \\ 
    $H_2^+ H_2^- H_2^0 $ & $ \tfrac{i[- 4 k^3 U_{12}^H \alpha - 4 v_L v_R^2 U_{22}^H \alpha - 4 v_R^3 U_{32}^H \alpha - 2 k v_R^2 U_{12}^H  \lambda_1 - 2 k^2 v_R U_{32}^H \lambda_3 - 2 k^2 v_L U_{22}^H \lambda_4 + \sqrt{2} k v_R U_{22}^H \kappa]}{(k^2 + v_R^2)}$ \\ 
    $H_2^+ H_2^- H_3^0 $ & $ \tfrac{i[- 4 k^3 U_{13}^H \alpha - 4 v_L v_R^2 U_{23}^H \alpha - 4 v_R^3 U_{33}^H \alpha - 2 k v_R^2 U_{13}^H  \lambda_1 - 2 k^2 v_R U_{33}^H \lambda_3 - 2 k^2 v_L U_{23}^H \lambda_4 + \sqrt{2} k v_R U_{23}^H \kappa]}{(k^2 + v_R^2)}$ \\ 
    $H_i^+ H_i^- A_2^0 $ & 0 \\ 
  \end{tabular}
  \caption{SSS couplings appearing in the two-loop contributions to $\Delta a_\mu$.\label{tab:2loop2}}
\end{table}

\begin{table}
  \centering\setlength\tabcolsep{5pt}\renewcommand{\arraystretch}{1.1}
  \begin{tabular}{cc}
    Interaction & Coupling\\ \hline
    $W_{L}^+ W_{L}^- h $ & $ i\tfrac{g_{L}^2}{2} [k U^H_{11} + v_L U^H_{21} ]$\\
    $W_{L}^+ W_{L}^- H_2^0 $ & $i\tfrac{g_{L}^2}{2} [k U^H_{12} + v_L U^H_{22} ]$\\
    $W_{L}^+ W_{L}^- H_3^0 $ & $ i\tfrac{g_{L,R}^2}{2} [k U^H_{13} + v_L U^H_{23} ]$\\
    $W_{L}^+ W_{L}^- A_2^0 $ & 0\\
    $W_{R}^+ W_{R}^- h $ & $ i\tfrac{g_{R}^2}{2} [k U^H_{11} + v_R U^H_{31} ]$\\
    $W_{R}^+ W_{R}^- H_2^0 $ & $i\tfrac{g_{R}^2}{2} [k U^H_{12} + v_R U^H_{32} ]$\\
    $W_{R}^+ W_{R}^- H_3^0 $ & $i\tfrac{g_{R}^2}{2} [k U^H_{13} + v_R U^H_{33} ]$\\
    $W_{R}^+ W_{R}^- A_2^0 $ & 0\\
  \end{tabular}\hspace{1.5cm}
  \begin{tabular}{cc}
    Interaction & Couplings\\ \hline
    $W_L^- H_1^+ h$ & $ -i \tfrac{g_L}{4} \tfrac{v_L U^H_{11} + k U^H_{21}}{ \sqrt{k^2+v_L^2}}$\\
    $W_L^- H_1^+ H_2^0$ & $ -i \tfrac{g_L}{4} \tfrac{v_L U^H_{12} + k U^H_{22}}{\sqrt{k^2+v_L^2}}$\\
    $W_L^- H_1^+ H_3^0$ & $ -i \tfrac{g_L}{4} \tfrac{v_L U^H_{13} + k U^H_{23}}{\sqrt{k^2+v_L^2}}$\\
    $W_L^- H_1^+ A_2^0$ & $ \tfrac{g_L}{ \sqrt{2}} U^A_{22}$\\
    $W_R^- H_2^+ h$ & $ -i \tfrac{g_R}{2} \tfrac{v_R U^H_{11} - k U^H_{31}}{ \sqrt{k^2+v_R^2}}$ \\
    $W_R^- H_2^+ H_2^0$ & $ -i \tfrac{g_R}{2} \tfrac{v_R U^H_{12} - k U^H_{32}}{\sqrt{k^2+v_R^2}}$ \\
    $W_R^- H_2^+ H_3^0$ & $ -i \tfrac{g_R}{2} \tfrac{v_R U^H_{13} - k U^H_{33}}{\sqrt{k^2+v_R^2}}$ \\
    $W_R^- H_2^+ A_2^0$ & $ - \tfrac{g_R}{ \sqrt{2}} U^A_{22}$ \\
  \end{tabular}    
  \caption{VVS and SSV couplings appearing in the two-loop contributions to $\Delta a_\mu$.\label{tab:2loop3}}
\end{table}

\begin{table}
  \centering\setlength\tabcolsep{4pt}\renewcommand{\arraystretch}{1.2}
  \resizebox{.98\columnwidth}{!}{\begin{tabular}{ccc}
    Interaction & Scalar ($g_S$) & Pseudo-scalar ($g_P$) \\ \hline
    $\overline{t} b H_1^+$ & $ \tfrac{i}{2} (V^\ast_{\rm CKM})_{33} \tfrac{v_L Y^q_{33} + k Y^q_{L,33}}{\sqrt{k^2 + v_L^2}} $ & $ - \tfrac{i}{2} (V^\ast_{\rm CKM})_{33} \tfrac{v_L Y^q_{33} - k Y^q_{L,33}}{\sqrt{k^2 + v_L^2}} $ \\
    $\overline{t} q_{d,i}^\prime H_2^+$ &  $ \tfrac{i}{2} (V^{\prime \ast}_{\rm CKM})_{3i} \tfrac{v_R Y_t + k Y^q_{R,ii}}{ \sqrt{k^2 + v_R^2}}$ &  $ - \tfrac{i}{2} (V^{\prime \ast}_{\rm CKM})_{31} \tfrac{v_R Y^q_{33} - k Y^q_{R,ii}}{ \sqrt{k^2 + v_R^2}}$ \\
  \end{tabular}\hspace{.5cm}
  \begin{tabular}{ccc}
    Interaction & Vector $(g_V)$ & Axial-vector $(g_A)$ \\ \hline
    $\overline{t} b W_L^+$ & $ i\tfrac{g_L}{2\sqrt{2}} (V^\ast_{\rm CKM})_{33}  $ & $  -i\tfrac{g_L}{2\sqrt{2}}  (V^\ast_{\rm CKM})_{33}  $ \\
    $\overline{t} q_{d,i}^\prime W_R^+$ &  $ i\tfrac{g_R}{2\sqrt{2}}  (V^{\prime \ast}_{\rm CKM})_{3i} $ &  $ i\tfrac{g_R}{2\sqrt{2}}  (V^{\prime \ast}_{\rm CKM})_{3i} $ \\[.07cm]
  \end{tabular}}
  \caption{Charged-current FFV and FFS couplings appearing in the two-loop contributions to $\Delta a_\mu$ of diagrams (6a) and (6b) in Fig.~\ref{fig:2loop1}, split in their scalar/vector and pseudo-scalar/axial-vector components. The index $i$ is a generation index. \label{tab:2loop6}}
\end{table}

\bibliographystyle{utcaps_mod}
\bibliography{ALRMlfv}
\end{document}